\documentstyle[11pt,rotate,epsf,psfig]{article} 

\newcommand{\be}{\begin{equation}}
\newcommand{\ee}{\end{equation}}
\newcommand{\bea}{\begin{eqnarray}}
\newcommand{\eea}{\end{eqnarray}}
\renewcommand{\theequation}{\arabic{section}.\arabic{equation}}
\newcommand{\lbl}[1]{\label{eq:#1}}
\newcommand{ \rf}[1]{(\ref{eq:#1})}

\newcommand{\FF}{{\cal F}_{\pi^0\gamma^*\gamma^*}}

\newcommand{\setl}{\setlength\arraycolsep{2pt}}

\newcommand{\noi}{\noindent}
\newcommand{\nn}{\nonumber}
\newcommand{\ra}{\rightarrow}

\newcommand{\cO}{{\cal O}}

\newcommand{\GeV}{\mbox{\rm GeV}}

\newcommand{\GF}{G_{\mbox{\rm {\tiny F}}}}

\newcommand{\lapprox}{%
\mathrel{%
\setbox0=\hbox{$<$}
\raise0.6ex\copy0\kern-\wd0
\lower0.65ex\hbox{$\sim$}
}}
\newcommand{\gapprox}{%
\mathrel{%
\setbox0=\hbox{$>$}
\raise0.6ex\copy0\kern-\wd0
\lower0.65ex\hbox{$\sim$}
}}

\begin{document}
\begin{flushright}
\today  \\
CPT-2003/P.4525\\
\end{flushright}

\vspace*{0.2cm}
\begin{center}
{\Large {\bf 
The anomalous magnetic moment of the muon:
a theoretical introduction~\footnote
{Based on the lectures delivered at the
41. Internationale Universitätswochen für Theoretische Physik,
Schladming, Styria, Austria, February 22 - 28, 2003.}
}}\\[0.8cm]
Marc Knecht \\[0.3cm]
Centre de Physique Th\'{e}orique, CNRS-Luminy, Case 907\\ 
    F-13288 Marseille Cedex 9, France

\indent

\indent

\indent

%
\end{center}
\noindent

\newpage

\section{Introduction}
\renewcommand{\theequation}{\arabic{section}.\arabic{equation}}
\setcounter{equation}{0}

In February 2001, the Muon (g-2) Collaboration of the E821 experiment 
at the Brookhaven AGS released a new value of the anomalous 
magnetic moment of the muon $a_\mu$,
measured with an unprecedented accuracy of 1.3 ppm [parts per million].
This annoucement has caused quite some excitement in the 
particle physics community. Indeed, this experimental
value was claimed to show a deviation of 2.6 $\sigma$ with one of 
the most accurate evaluations of the anomalous 
magnetic moment of the muon within the standard model. It was
subsequently shown that a sign error in one of the theoretical
contributions was responsible for a sizeable part of this discrepancy,
which eventually only amounted to 1.6 $\sigma$.
However, this event had the merit to draw the attention to the fact
that low energy but high precision experiments represent real 
potentialities,
complementary to the high energy accelerator programs,
for evidencing possible new degrees of freedom, 
supersymmetry or whatever else,
beyond those described by the standard model of electromagnetic, weak, 
and strong interactions.

\indent

\noindent
Clearly, in order for theory to match such an accurate measurement, 
calculations in the standard model have to be pushed
to their very limits. The difficulty is not only one of having to 
compute higher orders in perturbation theory, but also to correctly
take into account strong interaction contributions involving low-energy 
scales, where non perturbative effects are important, and which therefore
represent a real theoretical challenge. 
Furthermore, in the meantime the members of experimental team at 
Brookhaven have further improved upon their measurement. In July 2002,
they have announced a new result, in perfect agreement with the one they
had obtained about one year and a half before, but with an error
lowered down to the 0.7 ppm level. In addition, the theoretical
evaluation of a specific and important contribution, called
hadronic vacuum polarization, on which we shall have more to say
later on, has shown a discrepancy between different data sets
that are used as imputs. Depending on the choice between these
conflicting data, the discrepancy between the experimental and
the theoretical values can be as small as 1 $\sigma$, certainly
not a case for beyond the standard model physics, or can reach
the 3 $\sigma$ level, a much more promising situation as far as
the possibility of ``new physics'' is concerned.

\indent

\noindent
The purpose of this account is to give an overview of the main
features of the theoretical calculations that have been done
in order to obtain accurate predictions for the anomalous
magnetic moment of the muon within the 
standard model. Actually, all three charged leptons, $e^\pm$,
$\mu^\pm$, and $\tau^\pm$, of the standard model can be treated
on the same footing, except that the very different values of their
masses will induce different sensitivities with respect to the mass
scales involved in the higher order quantum corrections. Thus, the 
anomalous magnetic moment of the electron is almost only [but not quite]
sensitive to the electromagnetic interactions of the leptons, and
its value is barely affected by strong interaction effects or by
weak interaction corrections. On the other hand, the strengths of the
latter two types of corrections are enhanced by a considerable factor,
$\sim (m_\tau/m_e)^2\sim 1.2\times 10^7$, in the case of the $\tau$,
as compared to the electron. The same huge enhancement factor would also
affect the contributions coming from degrees of freedom beyond the
standard model, so that the measurement of the anomalous magnetic
moment of the $\tau$ would represent the best opportunity to detect
new physics. Unfortunately, the very short lifetime of the $\tau$
lepton which, precisely because of its high mass, can also decay
into hadronic states, makes such a measurement impossible at
present. The muon lies somewhat in the intermediate
range of mass scales \footnote{The corresponding enhancement
factor is  $\sim (m_\mu/m_e)^2\sim 4\times 10^4$.} and its 
lifetime still makes a 
measurement of its anomalous magnetic moment possible.
However, in order to obtain an accurate [that is, 
comparable to the experimental accuracy] prediction, the contributions of all
the sectors of the standard model have to be known very precisely.
Therefore, although the other two lepton flavours will also be discussed, 
since this does not really require additional work, the emphasis 
of these lectures will nevertheless be put on the muon.  

\indent

\noindent
There exist several excellent reviews and introductions, which the 
interested reader may consult. As far as the situation up to 1990
is concerned, the collection of articles published in Ref. 
\cite{QED} offers a wealth of information, on both
theory and experiment. Very useful accounts of earlier theoretical
work are presented in Refs. \cite{LautrupdeRafael72,CalmetetalRMP77}. 
Among the more recent reviews, Refs. 
\cite{CzarneckiMarciano98,HughesKinoshita99,Melnikov01,EdeRLaThuille02,Nyffeler03}
are most informative. I shall not touch the subject of the study of
new physics scenarios which might offer an explanation for a
possible deviation between the standard model prediction of the
magnetic moment of the muon and its experimental value, should
such a deviation be confirmed in the future. For this aspect,
I refer the reader to \cite{CzarneckiMarciano01} and to the articles 
quoted therein, or to \cite{gminus2URL} for a list of the recent
papers on the subject.

\indent


\section{General considerations}
\renewcommand{\theequation}{\arabic{section}.\arabic{equation}}
\setcounter{equation}{0}

In the context of relativistic quantum mechanics, the interaction 
of a pointlike spin one-half particle of charge $e_{\ell}$ and mass 
$m_{\ell}$ [$\ell$ stands hereafter for
any of the three charged lepton flavours $e$, $\mu$ or $\tau$]
with an external electromagnetic
field ${\cal A}_{\mu}(x)$ is described by the Dirac equation with the 
minimal coupling prescription,
\be
i\hbar \,\frac{\partial\psi}{\partial t}\,=\,\left[
c{\mbox{\boldmath$\alpha$}}\cdot\left(-i\hbar{\mbox{\boldmath$\nabla$}} - 
\frac{e_{\ell}}{c}{\mbox{\boldmath$\cal A$}}\right)
+ \beta m_{\ell} c^2 + e_{\ell} {\cal A}_0 \right]\psi
\,.
\ee
In the non relativistic limit, this reduces to the Pauli equation for the two-component
spinor $\varphi$ describing the large components of the Dirac spinor $\psi$,
\be
i\hbar \,\frac{\partial\varphi}{\partial t}\,=\,\left[
\frac{(-i\hbar{\mbox{\boldmath$\nabla$}} - 
({e_{\ell}}/c){\mbox{\boldmath$\cal A$}})^2}{2m_{\ell}}\, - \,
\frac{e_{\ell}\hbar}{2m_{\ell}c}\,{\mbox{\boldmath$\sigma$}}\cdot{\mbox{\bf B}}\,+\,
e_{\ell}{\cal A}_0 \right]\varphi
\,.
\ee
As is well known, this equation amounts to associate with the particle's spin
a magnetic moment
\be
{\mbox{\bf M}}_{s} \,=\,g_{\ell}\,\left(\frac{e_{\ell}}{2m_{\ell}c}\right)\,{\mbox{\bf S}}\,,
\ {\mbox{\bf S}} = \hbar\,\frac{\mbox{\boldmath$\sigma$}}{2}
\,,
\ee
with a gyromagnetic ratio predicted to be $g_{\ell}=2$.

\indent

\noindent
In the context of quantum field theory, the response to an external electromagnetic 
field is described by the matrix element of the electromagnetic current
\footnote{In the standard model, ${\cal J}^{\rho}$ denotes the total
electromagnetic current, with the contributions of all the charged 
elementary fields in presence, 
leptons, quarks, electroweak gauge bosons,...}${\cal J}^{\rho}$
[spin projections and Dirac 
indices of the spinors are not written explicitly]
\be
\langle \ell^-(p\,')\vert {\cal J}^{\rho}(0)\vert \ell^-(p)\rangle \,=\,
{\bar{\mbox{u}}}(p\,')\Gamma^{\rho}(p\,',p){\mbox{u}}(p)\,,
\lbl{defGamma}
\ee
with [$k_{\mu}\equiv p_\mu ' - p_{\mu}$]
\be 
\Gamma^{\rho}(p\,',p)\,=\,F_1(k^2)\gamma^{\rho}\,+\,\frac{i}{2m_{\ell}}\,F_2(k^2)
\sigma^{\rho\nu}k_{\nu}
\,-\,F_3(k^2)\gamma_5\sigma^{\rho\nu}k_{\nu}
\,+\,F_4(k^2)[k^2\gamma^\rho - 2m_\ell k^\rho]\gamma_5
\,.
\lbl{defFis}
\ee
This expression of the matrix element 
$\langle \ell^-(p\,')\vert {\cal J}^{\rho}(0)\vert \ell^-(p)\rangle$ is the most general
that follows from Lorentz invariance, the Dirac equation for the two spinors,
$(\not\! p -m){\mbox{u}}(p) = 0$, ${\bar{\mbox{u}}}(p\,')(\not\!p\,' - m) =0$, and the 
conservation of the electromagnetic current, 
$\left(\partial\cdot{\cal J}\right)(x) = 0$.
The two first form factors, $F_1(k^2)$ and $F_2(k^2)$, are known as the
Dirac form factor and the Pauli form factor, respectively. 
Since the electric charge operator ${\cal Q}$ is given, in units of the charge $e_{\ell}$, by 
\be
{\cal Q} \,=\,\int d{\bf{\mbox{x}}}\, {\cal J}_0(x^0,{\bf{\mbox{x}}})
\,,
\ee
the form factor $F_1(k^2)$ is normalized by the condition $F_1(0) = 1$. 
The presence of the form factor $F_3(k^2)$ requires both parity and time
reversal invariance to be broken, whereas $F_4(k^2)$ can be different from
zero provided parity is broken. Both $F_3(k^2)$ and $F_4(k^2)$  are 
therefore absent if only electromagnetic and strong
interactions are considered [we leave aside the possibility of
having a non vanishing vacuum angle in the strong interaction sector]. 
On the other hand, in the standard model, the weak interactions
violate both parity and time reversal symmetry, so that they actually
induce non vanishing expressions for these form factors.

\indent

\noindent
The above form factors are defined for $k^2<0$, and they are
real in this region if the current ${\cal J}_\rho(x)$ is hermitian.
Due to  general properties of 
quantum field theory, like causality, analyticity, and crossing symmetry,
these form factors can be analytically continued into the whole
complex $k^2$ plane with a cut for $k^2>4m_\ell^2$. They then 
become complex functions, obeying the Schwartz reflection
property $F_i(k^2)^\ast = F_i(k^{2\ast})$. For $k^2>4m_{\ell}^2$,
the form factors $F_i(k^2+i\epsilon)$ describe the crossed
channel matrix element
$\langle \ell^-(p\,') \ell^+(p)\vert {\cal J}^{\rho}(0)\vert 0\rangle$.
Furthermore, at $k^2=0$, they describe the residue of the s-channel 
photon pole in the S-matrix element for elastic $\ell^+\ell^-$ scattering.

\indent

\noindent
At tree level in the standard model, one finds
\be
F_1^{\,\scriptsize{\mbox{tree}}}(k^2)=1
\,,\ F_i^{\,\scriptsize{\mbox{tree}}}(k^2)=0,\,i=2,\,3,\,4
\,.
\ee
In order to obtain non zero values for $F_2(k^2)$, $F_3(k^2)$, and $F_4(k^2)$ already
at tree level, the interaction of the Dirac field with the photon field ${\cal A}_\mu$
would have to depart from the minimal coupling prescription. For instance, the modification 
[${\cal F}_{\mu\nu}=\partial_\mu {\cal A}_\nu - \partial_\nu {\cal A}_\mu$,
${\cal J}^\rho = {\overline\psi}\gamma^\rho\psi$]
\bea
\int d^4x {\cal L}_{\scriptsize{\mbox{int}}} \,=\, 
-\frac{e_{\ell}}{c}\int d^4x {\cal J}^{\rho}{\cal A}_{\rho}
\ \rightarrow 
\nonumber\\
\!\!\!\!\!\!\!\!\!\!\!\!\!\!\!\!\!\!\!\!
\rightarrow
\ \int d^4x {\widehat{\cal L}}_{\scriptsize{\mbox{int}}} &=& -\frac{e_{\ell}}{c}
\int d^4x \bigg[
{\cal J}^{\rho}{\cal A}_{\rho}
+ \frac{\hbar}{4m_{\ell}}a_{\ell}{\overline\psi}\sigma_{\mu\nu}{\psi}{\cal F}^{\mu\nu}
+ \frac{\hbar}{2e_{\ell}}d_{\ell}{\overline\psi}i\gamma_5\sigma_{\mu\nu}{\psi}{\cal F}^{\mu\nu}
\bigg]
\nonumber\\
&=&
-\frac{e_{\ell}}{c}\int d^4x{\widehat{\cal J}}^{\rho}{\cal A}_{\rho}
\,,
\lbl{Lintmod}
\eea
with \footnote{The current ${\widehat{\cal J}}^{\rho}$ is still a conserved four-vector,
therefore the matrix element 
$\langle \ell^-(p\,')\vert {\widehat{\cal J}}^{\rho}(0)\vert \ell^-(p)\rangle$ also takes the 
form \rf{defGamma}, \rf{defFis}, with appropriate form factors ${\widehat F}_i(k^2)$.}
\be
{\widehat{\cal J}}_{\rho} \,=\, {\cal J}_{\rho} 
- \frac{\hbar}{2m_{\ell}}a_{\ell}
\partial^\mu\big({\overline\psi}\sigma_{\mu\rho}{\psi}\big)
- \frac{\hbar d_\ell}{e_\ell}
\partial^\mu\big({\overline\psi}i\gamma_5\sigma_{\mu\rho}{\psi}\big)
\,,
\ee
leads to
\be
{\widehat F}_1^{\,\scriptsize{\mbox{tree}}}(k^2)=1
\,,\ {\widehat F}_2^{\,\scriptsize{\mbox{tree}}}(k^2)=a_{\ell}
\,,\ {\widehat F}_3^{\,\scriptsize{\mbox{tree}}}(k^2)=d_{\ell}/e_{\ell}
\,,\ {\widehat F}_4^{\,\scriptsize{\mbox{tree}}}(k^2)=0\,.
\ee
The equation satisfied by the Dirac spinor $\psi$ then reads
\bea
&&
\!\!\!\!\!\!\!\!\!\!\!\!\!\!\!\!\!\!\!\!\!\!\!\!\!\!\!\!\!\!\!\!\!\!\!\!\!\!\!\!
i\hbar \,\frac{\partial\psi}{\partial t}\,=\,\bigg[
c{\mbox{\boldmath$\alpha$}}\cdot\left(-i\hbar{\mbox{\boldmath$\nabla$}} - 
\frac{e_{\ell}}{c}{\mbox{\boldmath$\cal A$}}\right)
+ \beta m_{\ell} c^2 + e_{\ell} {\cal A}_0 
\nonumber\\
&&\qquad
+ \frac{e_{\ell}\hbar}{2m_{\ell}}\,a_{\ell}\beta
\left(i{\mbox{\boldmath$\alpha$}}\cdot{\mbox{\bf E}} - {\mbox{\boldmath$\Sigma$}}\cdot{\mbox{\bf B}}\right)
- \hbar d_{\ell}\beta
\left({\mbox{\boldmath$\Sigma$}}\cdot {\mbox{\bf E}} + i{\mbox{\boldmath$\alpha$}}\cdot {\mbox{\bf B}}\right)
\bigg]\psi
\,,
\eea
and the corresponding non relativistic limit becomes \footnote{Terms involving the
gradients of the external fields {\bf E} and {\bf B} or terms nonlinear in these
fields are not shown.}
\be
i\hbar \,\frac{\partial\varphi}{\partial t}\,=\,\left[
\frac{(-i\hbar{\mbox{\boldmath$\nabla$}} - ({e_{\ell}}/c){\mbox{\boldmath$\cal A$}})^2}{2m_{\ell}}\, - \,
\frac{e_{\ell}\hbar}{2m_{\ell}c}\,(1+a_{\ell})
{\mbox{\boldmath$\sigma$}}\cdot{\mbox{\bf B}}
\,-\, \hbar d_{\ell} {\mbox{\boldmath$\sigma$}}\cdot{\mbox{\bf E}}
\,+\,e{_{\ell}\cal A}_0 \,+\, \cdots \right]\varphi
\,.
\ee
Thus the coupling constant $a_{\ell}$ induces a shift in the gyromagnetic factor,
$g_\ell = 2(1+a_\ell)$,
while $d_{\ell}$ gives rise to an electric dipole moment. 
The modification
\rf{Lintmod} of the interaction with the photon field introduces two arbitrary
constants, and both terms produce a {\em non renormalizable} interaction.
Non constant values of the form factors could be generated at tree level 
upon introducing \cite{Foldy52} additional non renormalizable couplings, involving 
derivatives of the external field of the type $\Box^n{\cal A}_{\mu}$, which
preserve the gauge invariance of the corresponding field equation satisfied by $\psi$.
In a similar way, one can also introduce terms which induce a nonzero
value for $F_4$.
In a renormalizable framework, like QED or the standard model,
calculable non vanishing values for $F_2(k^2)$, $F_3(k^2)$, 
and $F_4(k^2)$ are generated by the loop corrections. 
In particular, the latter will likewise induce 
an {\em anomalous magnetic moment} 
\be
a_{\ell}=\frac{1}{2}(g_\ell - 2)=F_2(0)
\ee 
and an electric dipole moment $d_{\ell}=e_{\ell}F_3(0)$;
$F_4(0)$, which corresponds to an axial radius of the lepton,
is also called the {\it anapole moment} \cite{Zeldovich58,Zeldovich61,Marshak},
and is sensitive to the gradients of the external fields.

\indent

\noindent
If we consider only the electromagnetic and the strong interactions, 
the current ${\cal J}^{\rho}$ is gauge invariant, and the two 
form factors that remain in that case, 
$F_1(k^2)$ and $F_2(k^2)$, do not depend on the gauges chosen in order
to quantize the photon and the gluon gauge fields. This is no longer the case if
the weak interactions are included as well, since ${\cal J}^{\rho}$ now transforms 
in a non trivial way
under a weak gauge transformation, and the corresponding form factors in 
general depend on the gauge choices. As we have already mentioned above,
the zero momentum transfer values
$F_i(0)$, $i=1,2,3,4$ describe a physical S-matrix element. 
To the extent that the perturbative S-matrix
of the standard model does not depend on the gauge fixing
parameters to any order of the renormalized perturbation
expansion, the quantities $F_i(0)$ should define {\it bona fide} 
gauge-fixing independent observables.

\indent

\noindent
The computation of $\Gamma_{\rho}(p\,',p)$ is often a tedious task, especially if
higher loop contributions are considered. It is therefore useful to concentrate
the efforts on computing the form factor of interest, e.g. $F_2(k^2)$ in the 
case of the anomalous magnetic moment. This can be achieved upon projecting 
out the different form factors \cite{BrodskySullivan67,Barbierietal72} 
using the following general expression~\footnote{From now on, I most of 
the time use the system of units where $\hbar =1$, $c = 1$. Other
projectors on $F_2(k^2)$ have also been devised, see e.g. 
\cite{Pietschmann64}, but are not currently used.}
\be
F_i(k^2)\,=\,{\mbox{tr}}\left[ \Lambda_i^{\rho}(p\,',p)(\not\!p\,'+m_{\ell})
\Gamma_{\rho}(p\,',p)(\not\!p+m_{\ell})\right]
\,,
\lbl{projF_i}
\ee
with
\bea
\Lambda_1^{\rho}(p\,',p) &=& \frac{1}{4}\,\frac{1}{k^2-4m^2_{\ell}}\,\gamma^{\rho}\,+
\,\frac{3m_{\ell}}{2}\,\frac{1}{(k^2-4m^2_{\ell})^2}\,(p\,'+p)^{\rho}
\nonumber\\
\Lambda_2^{\rho}(p\,',p) &=& -\,\frac{m^2_{\ell}}{k^2}\,
\frac{1}{k^2-4m^2_{\ell}}\,\gamma^{\rho}\,-
\,\frac{m_{\ell}}{k^2}\,\frac{k^2+2m^2_{\ell}}{(k^2-4m^2_{\ell})^2}\,(p\,'+p)^{\rho}
\nonumber\\
\Lambda_3^{\rho}(p\,',p) &=& -\,\frac{i}{2k^2}\,
\frac{1}{k^2-4m^2_{\ell}}\,\gamma_5(p\,'+p)^{\rho}
\nonumber\\
\Lambda_4^{\rho}(p\,',p) &=& -\,\frac{1}{4k^2}\,
\frac{1}{k^2-4m^2_{\ell}}\,\gamma_5\gamma^{\rho}
\,.
\lbl{projectors}
\eea

\noindent
For $k\to 0$, one has
\be
\Lambda^{\rho}_{2}(p\,',p) \ =\ \frac{1}{4k^2}\,\Big[\gamma^{\rho}\,-\,\frac{1}{m_{\ell}}\, \Big( 1+\frac{k^2}{m_{\ell}^2}\Big)(p+\frac{1}{2}k)^{\rho}\,+\,\cdots\Big]\,,
\lbl{proj1}
\ee
and
\be
 (\not\!p+m_{\ell})\Lambda^{\rho}_{2}(p\,',p) (\not\! p\,'+m_{\ell})\ = 
\ \frac{1}{4}\,(\not\! p +m_{\ell})\,\Big[-\,\frac{k^{\rho}}{k^2}\,+\,(\gamma^{\rho} \,-\, \frac{p^{\rho}}{m_{\ell}})\,\frac{\not\! k}{k^2}\,+\,\cdots\Big]\,.
\lbl{proj2}
\ee
The last expression behaves as $\sim 1/k$ as the external photon four 
momentum
$k_{\mu}$ vanishes, so that one may worry about the finiteness of $F_2(0)$
obtained upon using Eq. \rf{projF_i}. This problem is solved by the fact that
$\Gamma^{\rho}(p\,',p)$ satisfies the Ward identity
\be
(p\,' - p)_{\rho}\Gamma^{\rho}(p\,',p) \,=\, 0
\,,
\ee
following from the conservation of the electromagnetic current. 
Therefore, the identity
\be
\Gamma^{\rho}(p\,',p) \,=\, - k_{\sigma}\,\frac{\partial}{\partial k_{\rho}}
\,\Gamma^{\sigma}(p\,',p)
\ee
provides the additional power of $k$ which ensures a finite result as
$k_{\mu}\to 0$.

\indent

\noindent
The presence of three different interactions in the standard model naturally leads
one to consider the following decomposition of the anomalous magnetic moment $a_{\ell}$:
\be
a_{\ell} \,=\, a_{\ell}^{\scriptsize{\mbox{QED}}} \,+\, a_{\ell}^{\scriptsize{\mbox{had}}}
\,+\, a_{\ell}^{\scriptsize{\mbox{weak}}}
\,.
\ee
The first term, $a_{\ell}^{\scriptsize{\mbox{QED}}}$, denotes all the contributions which arise from loops
involving only virtual photons and leptons. Among these, it is useful to distinguish those
which involve only the same lepton flavour $\ell$ for which we wish to compute the anomalous
magnetic moment, and those which involve loops with leptons of different flavours, denoted
collectively as $\ell\,'$ [$\alpha\equiv e^2/4\pi$],
\be
a_{\ell}^{\scriptsize{\mbox{QED}}} \,=\, \sum_{n\ge 1} A_n\,\left(\frac{\alpha}{\pi}\right)^n \,+\,
\sum_{n\ge 2} B_n(\ell, \ell ')\,\left(\frac{\alpha}{\pi}\right)^n
\,.
\lbl{a_lQED}
\ee
The second type of contribution, $a_{\ell}^{\scriptsize{\mbox{had}}}$, 
involves also quark loops. Their contribution
is far from being limited to the short distance scales, and $a_{\ell}^{\scriptsize{\mbox{had}}}$
is an intrinsically non perturbative quantity. From a theoretical
point of view, this represents a serious difficulty. Finally, at some level of precision, the weak
interactions can no longer be ignored, and contributions of virtual Higgs or massive
gauge boson degrees of freedom induce the third component $a_{\ell}^{\scriptsize{\mbox{weak}}}$.
Of course, starting from the two loop level, a hadronic contribution to
$a_{\ell}^{\scriptsize{\mbox{weak}}}$ will also be present. 
The remainder of this 
presentation is devoted to a detailed discussion of these various contributions.

\indent

\noindent
Before starting this guided tour of the anomalous magnetic
moments of the massive charged leptons of the standard model, it is useful to
keep in mind a few simple considerations:

\indent

\noindent
$\bullet$ The anomalous magnetic moment is a dimensionless quantity. 
Therefore, the coefficients
$A_n$ above are {\it universal}, i.e. they do not depend on the flavour 
of the lepton whose
anomalous magnetic moment we wish to evaluate.

\indent

\noindent
$\bullet$ The contributions to $a_{\ell}$ of degrees of freedom corresponding to a 
typical scale $M\gg m_{\ell}$ decouple \cite{AppelquistCarazzone75}, i.e. they are 
{\it suppressed} by powers of $m_{\ell}/M$.\footnote{In the presence of the
weak interactions, this statement has to be reconsidered. Indeed, the necessity for
the cancellation of the $SU(2)\times U(1)$ gauge anomalies 
\cite{BIM72,GrossJackiw72,KorthalsPerrottet72} transforms 
the decoupling of, say, a single heavy fermion in a given generation,
into a somewhat subtle issue \cite{VeltmanSterling81,dHokerFarhi84}, 
the resulting lagrangian being no longer renormalizable.}

\indent

\noindent
$\bullet$ The contributions to $a_{\ell}$ originating from light degrees of freedom, characterized
by a typical scale $m\ll m_{\ell}$ are {\it enhanced} by powers of $\ln(m_{\ell}/m)$. 
At a given order, the logarithmic terms that do not vanish as  $m_{\ell}/m\to 0$ 
can often be computed from the
knowledge of the lower order terms and of the $\beta$ function through the 
renormalization group equations 
\cite{Kinoshita67,LautrupdeRafael74,deRafaelRosner74,KinoshitaMarcianoQED}.

\indent

\noindent
These general properties already allow to draw several elementary conclusions. 
The electron being the lightest charged lepton, its anomalous magnetic moment 
is dominantly determined by the values of the coefficients $A_n$. 
The first contribution of other degrees of freedom comes from graphs
involving, say, at least one muon loop, which occurs first at the two-loop 
level, and is of the order of $(m_{e}/m_{\mu})^2(\alpha/\pi)^2\sim 10^{-10}$. 
The hadronic effects, i.e. ``quark and gluon loops'', 
characterized by a scale of $\sim 1$ GeV, or effects of degrees of freedom beyond the 
standard model, which may appear at some high scale $M$, will be felt more strongly, 
by a considerable factor $(m_{\mu}/m_e)^2\sim 40\,000$, in $a_{\mu}$ than in $a_e$. 
Thus, $a_e$ is well suited for testing the validity of QED at higher orders, whereas 
$a_{\mu}$ is more appropriate  for testing the weak sector of the
standard model, one of the main motivations for the BNL experiment,
and possibly for detecting new physics.

\indent

\begin{center}
{\bf
Exercises for section 2
}
\end{center}

\noindent
{\it Exercise 2.1}\\
Show that the expression of the matrix element
of the electromagnetic current given by
 Eqs. \rf{defGamma} and \rf{defFis} indeed follows 
from the conditions stated. Show that the
form factors $F_i(k^2)$ in these equations
are real if the current ${\cal J}^{\rho}$ 
is hermitian. Work out the transformation
properties of the different form factors
under the operations of parity and time
reversal. How many additional
form factors are needed in order to describe the 
same matrix element if the assumption concerning the
conservation of the current is dropped?\\
\\
{\it Exercise 2.2}\\
Show that the current ${\widehat{\cal J}}^{\rho}$ 
defined by Eq. \rf{Lintmod} is conserved.\\
\\
{\it Exercise 2.3}\\
Find the term one needs to add to
${\widehat{\cal L}}_{\scriptsize{\mbox{int}}}$,
and thus to ${\widehat{\cal J}}_{\rho}$,
such as to generate a constant but nonzero 
form factor ${\widehat{F}}_4$ at tree level.
Show that in the non relativistic limit it
induces an interaction term of the form
$F_4(0){\mbox{\boldmath$\sigma$}}\cdot
({\mbox{\boldmath$\nabla$}}\wedge{\mbox{\bf B}})$
in a non uniform magnetic field.\\
\\
{\it Exercise 2.4}\\
Show that the quantities $\Lambda_i^\rho(p\,',p)$
defined in Eq. \rf{projectors}
indeed project on the corresponding form factors 
$F_i(k^2)$ through Eq. \rf{projF_i}.
Derive Eqs. \rf{proj1} and \rf{proj2}.\\
\\
{\it Exercise 2.5}\\
Work out the expression of the electromagnetic
current in the standard model.\\
\\
{\it Exercise 2.6}\\
Give the most general decomposition of the matrix element
in Eq. \rf{defFis} in the case of a massive Majorana
neutrino [hint: see Refs. \cite{Kayser82, MohapatraPal} for a 
rather complete treatment].

\indent

\section{Brief overview of the experimental situation}
\renewcommand{\theequation}{\arabic{section}.\arabic{equation}}
\setcounter{equation}{0}

\subsection{Measurements of the magnetic moment of the electron}

The first indication that the gyromagnetic factor of the electron is
different from the value $g_e = 2$ predicted by the Dirac theory came from
the precision measurement of hyperfine splitting in hydrogen and 
deuterium \cite{NafeNelsonRabi47}. The first measurement of
the gyromagnetic factor of free electrons was performed in 1958
\cite{Dehmelt58}, with
a precision of 3.6\%. The situation began to improve with the introduction of
experimental setups based on the Penning  trap. 
Some of the successive values obtained over a period of forty years
are shown in Table \ref{tab:tab1}. Technical improvements, eventually 
allowing for the trapping of a single electron or positron, produced,
in the course of time,
an enormous increase in precision which, starting from a few percents, 
went through the ppm 
levels, before culminating at 4 ppb [parts per billion] 
 in the last \cite{VanDycketal87} of a series 
of experiments performed at the University of Washington in Seattle.
The same experiment has also
produced a measurement of the magnetic moment 
of the positron with the same accuracy, thus providing a test of
$CPT$ invariance at the level of $10^{-12}$,
\be
g_{e^-}/g_{e^+} \,=\, 1 \,+\,(0.5\pm2.1)\times 10^{-12}
\,.
\ee
Assuming invariance under $CPT$, the weighted average
of the electron and positron anomalous moments obtained in
Ref. \cite{VanDycketal87} gives \cite{MohrTaylorRMP00,PDG02},
\be
a_e^{\mbox{\scriptsize{exp}}}\,=\,0.001\,159\,652\,188\,3(4\,2) 
\,.
\lbl{a_eexpaverage}
\ee
An extensive survey of the literature and a detailed description of the 
various experimental aspects can be found in \cite{VanDyckQED}. The 
earlier experiments are reviewed in \cite{RichWesley72}.

\begin{table}[!h]
\caption{Some experimental determinations of the electron's 
anomalous magnetic moment $a_e$ with the corresponding relative
precision.}
\begin{center}
\renewcommand{\arraystretch}{1.1}
\begin{tabular}{lcc}
\hline\hline
0.001\,19(5)   &     4.2\%           &      \protect{\cite{KuschFowley47}}  \\
0.001\,165(11) &   1\%             &      \protect{\cite{FrankenLiebes56}}  \\
0.001\,116(40) &   3.6\%             &      \protect{\cite{Dehmelt58}}   \\
0.001\,160\,9(2\,4)  &   2\,100 {\mbox ppm} & \protect{\cite{Schuppetal61}}\\
0.001\,159\,622(27) &       23 {\mbox ppm}     & \protect{\cite{WilkinsonCrane63}}\\
0.001\,159\,660(300) & 258 {\mbox ppm} & \protect{\cite{Graeffetal68}}\\
0.001\,159\,657\,7(3\,5) & 3 {\mbox ppm} & \protect{\cite{WesleyRich71}}\\
0.001\,159\,652\,41(20) & 172 {\mbox ppb} & \protect{\cite{VanDycketal77}}\\
0.001\,159\,652\,188\,4(4\,3) & 4 {\mbox ppb} & \protect{\cite{VanDycketal87}}\\
\hline\hline
\end{tabular}
\label{tab:tab1} 
\end{center}
\end{table}

\indent

\subsection{Measurements of the magnetic moment of the muon}

The anomalous magnetic moment of the muon has also
been the subject of quite a few experiments. The
very short lifetime of the muon, 
$\tau_\mu = (2.19703\pm 0.00004)\times 10^{-6} s$,
makes it necessary to proceed in a completely
different way in order to attain a high precision. 
The experiments conducted at CERN during the 
years 1968-1977 used a muon storage ring [for details,
see \cite{FarleyPicassoQED} and references quoted therein].
The more recent experiments at the AGS in Brookhaven
are based on the same concept.
Pions are produced by sending a proton beam 
on a target. The pions subsequently 
decay into longitudinally polarized muons, which are 
captured inside a storage ring, where they follow a 
circular orbit in the presence of both a uniform magnetic field
and a quadrupole electric field, the latter serving the purpose
of stabilizing the orbits.
The difference between the spin precession frequency 
and the orbit frequency is given by
\be
{\mbox{\boldmath$\omega$}}_s \,-\, {\mbox{\boldmath$\omega$}}_c 
\,=\, -\,\frac{e}{m_\mu c}\,\left\{
a_\mu {\mbox{\bf B}} \,-\, \left[a_\mu\,+\,\frac{1}{1-\gamma^2} \right]
{\mbox{\boldmath$\beta$}}\wedge {\mbox{\bf E}}\right\}
\,,
\lbl{precess1}
\ee
where ${\mbox{\boldmath$\beta$}}$ is the velocity of the muons,
and $\gamma$ is the corresponding Lorentz boost factor.
Therefore, if $\gamma$ is tuned to its
``magic'' value $\gamma = \sqrt{1+1/a_\mu} = 29.3$,
the measurement of $\omega_s - \omega_c $ and of the
magnetic field {\bf B} allows to determine $a_\mu$. The spin direction of the
muon is determined by detecting the 
electrons or positrons produced in the decay
of the muons with an energy greater than some threshold energy $E_t$.
The number of detected electrons $N_e(t)$ decreases exponentially with 
time, the time constant being set by the muon's lifetime $\gamma\tau_\mu$c 
in the laboratory frame, and is modulated by the frequency 
$\omega_s - \omega_c$,
\be
N_e(t)\,=\,N_0(E_t)e^{-t/\gamma\tau_\mu} 
\{1 + A(E_t)\cos[(\omega_s - \omega_c)t+\phi(E_t)]\}
\,.
\ee
The observation of this time dependence thus provides
the required measurement of $\omega_s - \omega_c$.

\begin{table}[!h]
\caption{Determinations of the  anomalous magnetic
moment of the positively charged muon from the storage ring
experiments conducted at the CERN PS 
and at the BNL AGS.}
\begin{center}
\renewcommand{\arraystretch}{1.1}
\begin{tabular}{lcc}
\hline\hline
0.001\,166\,16(31)    &  265 {\mbox ppm}   & \protect{\cite{Baileyetal68}}\\
0.001\,165\,895(27)    &  23 {\mbox ppm}   & \protect{\cite{Baileyetal75}}\\
0.001\,165\,911(11)   &  10 {\mbox ppm}    & \protect{\cite{Baileyetal79}}\\
0.001\,165\,925(15)   &  13 {\mbox ppm}    & \protect{\cite{Careyetal99}}\\
0.001\,165\,919\,1(5\,9)   &    5 {\mbox ppm}    & \protect{\cite{Brownetal00}}\\
0.001\,165\,920\,2(1\,6)   & 1.3 {\mbox ppm}  & \protect{\cite{Brownetal01}}\\
0.001\,165\,920\,3(8)   & 0.7 {\mbox ppm}  & \protect{\cite{Bennettetal02}}\\
\hline\hline
\end{tabular}
\label{tab:tab2} 
\end{center}
\end{table}

\indent

\noindent
Several experimental results for the anomalous magnetic
moment of the positively charged muon, obtained at the CERN PS
or, more recently, at the BNL AGS, are recorded in Table \ref{tab:tab2}.
Notice that the relative errors are measured in ppm units, 
to be contrasted with the ppb level of accuracy achieved in 
the electron case.
The four last values in Table \ref{tab:tab2} were 
obtained by the E821 experiment at BNL. They 
show a remarkable stability
and a steady increase in precision, and now completely
dominate the world average value. Further data, for negatively
charged muons \footnote{The CERN experiment had
also measured $a_{\mu^-}=0.001\,165\,937(12)$ with a 10 ppm 
accuracy, giving the average value $a_\mu=0.001\,165\,924(8.5)$, with
an accuracy of 7 ppm.} are presently being analysed.
The aim of
the Brookhaven Muon (g - 2) Collaboration  is to reach a precision of 0.35 ppm, 
but this will depend on whether the experiment will receive 
financial support to collect more data or not~\footnote{At
the time of writing, the prospects in this respect are unfortunately 
rather dim.}.

\indent

\noindent
For completeness, one should mention that Eq. \rf{precess1}
is only correct as long as the muon has no electric dipole
moment. If this is not the case, the more general relation,
\be
{\mbox{\boldmath$\omega$}}_s \,-\, {\mbox{\boldmath$\omega$}}_c 
\,=\, -\,\frac{e}{m_\mu c}\,\left\{
a_\mu {\mbox{\bf B}} \,-\, \left[a_\mu\,+\,\frac{1}{1-\gamma^2} \right]
{\mbox{\boldmath$\beta$}}\wedge {\mbox{\bf E}}\right\}\,-\,
\frac{2d_\mu}{\hbar}\,\left\{
{\mbox{\boldmath$\beta$}}\wedge{\mbox{\bf B}} \,+\, {\mbox{\bf E}}\right\}
\,,
\lbl{precess2}
\ee
which holds for ${\mbox{\boldmath$\beta$}}\cdot{\mbox{\bf E}}
={\mbox{\boldmath$\beta$}}\cdot{\mbox{\bf B}}=0$,
has to be used. The additional term proportional to $d_\mu$ induces an
oscillation of the muon spin with respect to the plane of 
motion~\footnote{A measurement of the electric dipole moment of 
the muon
was actually performed by the CERN experiment \cite{Baileyetal79}, with
the result $d_\mu = (3.7\pm 3.4)\times 10^{-19}\,e$cm. An even smaller
value, $d_\mu \lapprox 9.1 \times 10^{-25}\, e$cm, can be infered from 
the experimental value of the electric dipole moment of the electron 
\cite{Comminsetal94,Comminsetal02}, 
$d_e = 1.8(1.2)(1.0)\times 10^{-27},e$cm, and assuming
a scaling law $d_\mu \sim \frac{m_\mu}{m_e}d_e$. Such a scaling law holds 
within the standard model, but not in models with flavour violating
interactions, see for instance \cite{Fengetal}. For a proposal to measure 
$d_\mu$ at the level of $\sim 10^{-24},e$cm, see \cite{Semertzidis01}.}. 
In the standard model, given the experimental precision and the 
intensities of the fields used in the experiment, it is quite legitimate to
use the formula \rf{precess1} instead of \rf{precess2}. However, in case
a discrepancy arises between the experimental value of $a_\mu$ and the
standard model prediction, the difference could be induced by non standard
contributions to either the anomalous magnetic moment $a_\mu$ or 
the electric dipole moment $d_\mu$, see the discussion in \cite{Fengetal}.

\indent

\subsection{Experimental bounds on the anomalous magnetic moment of the 
$\tau$ lepton}

As already mentioned, the very short lifetime of the $\tau$
precludes a measurement of its anomalous magnetic moment following
any of the techniques described above. Indirect access to $a_\tau$
is provided by the reaction $e^+e^- \to \tau^+\tau^-\gamma$. The results
obtained by OPAL \cite{OPAL} and L3 \cite{L3} at LEP only lead to very 
loose bounds,
\bea
&&
-0.052 < a_\tau < 0.058\ (95\% C.L.)
\nonumber\\
&&
-0.068 < a_\tau < 0.065\ (95\% C.L.)
\,,
\eea
respectively.

\indent

\noindent
We shall now turn towards theory, in order to
see how the standard model predictions compare
with these experimental values.
Only the cases of the electron and of the muon will
be treated in some detail. The theoretical aspects as
far as the anomalous magnetic moment of the $\tau$ are
concerned are discussed in \cite{Narison01} and in the references
quoted therein.

\indent

\section{The anomalous magnetic moment of the electron}
\renewcommand{\theequation}{\arabic{section}.\arabic{equation}}
\setcounter{equation}{0}
 
\begin{figure}[!h]
\centerline{\psfig{figure=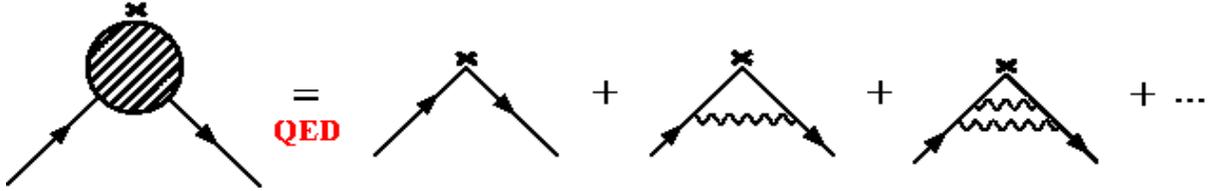,height=2.5cm,width=16cm}}
\caption{The perturbative expansion of $\Gamma^\rho(p\,',p)$ in single flavour QED. 
The tree graph gives $F_1=1$, $F_2=0$, whereas $F_3(k^2)$
and $F_4(k^2)$ vanish identically to all orders in pure QED. 
The one loop vertex correction graph gives the coefficient
$A_1$ in Eq. {\protect\rf{a_lQED}}. The cross denotes the 
insertion of the external field.} \label{fig:QED1loop}
\end{figure}

We start with the anomalous magnetic moment of the lightest charged lepton,
the electron. Since the electron mass $m_e$ is much smaller than any other mass
scale present in the standard model, the mass independent part of 
$a_e^{\scriptsize{\mbox{QED}}}$ dominates its value. As mentioned before, 
non vanishing contributions appear at the level of the loop diagrams
shown in Fig.~\ref{fig:QED1loop}. 

\indent

\subsection{The lowest order contribution}

The one loop diagram gives
\bea
\Gamma^{\rho}(p\,',p)\big\vert_{\scriptsize{\mbox{1 loop}}}  &=& 
(-ie)^2\int\frac{d^4q}{(2\pi)^4}
\gamma^{\mu}\,\frac{i}{\not\!p\,' + \not\!q - m_e}\,\gamma^{\rho}
\,\frac{i}{\not\!p \,+ \not\!q - m_e}\,\gamma^{\nu}
\nonumber\\
&&\qquad\times
\,\frac{(-i)}{q^2}\left[
\eta_{\mu\nu} \,-\,(1-\xi)\frac{q_\mu q_nu}{q^2}\right]
\,.
\lbl{GammaFig1}
\eea
The photon propagator has been written in a Lorentz type gauge,
corresponding to a covariant gauge fixing term 
$-(\partial\cdot A)^2/2\xi$. Let us for a moment concentate
on the $\xi$ - dependence of this one loop expression. Recall
that $\Gamma^{\rho}(p\,',p)$ has eventually to be inserted
between the spinors ${\bar{\mbox{u}}}(p\,')$ and 
${{\mbox{u}}}(p)$. Then, the gauge dependence of the
integrand is given by
\bea
(1 - \xi){\bar{\mbox{u}}}(p\,')\not\!q\,\frac{i}{\not\!p\,' + \not\!q - m_e}\,\gamma^{\rho}
\,\frac{i}{\not\!p\, + \not\!q - m_e}\,\not\!q {{\mbox{u}}}(p)
\,=\,(1 - \xi) {\bar{\mbox{u}}}(p\,') i \gamma_\rho i {{\mbox{u}}}(p)
\nonumber
\,,
\eea
and thus affects $F_1(k^2)$, but not $F_2(k^2)$. For evaluating
the latter, one may therefore take, say, $\xi =1$ for convenience.
The form factor $F_2(k^2)$ is then obtained by using Eqs. \rf{projF_i} and
\rf{projectors} and, upon evaluating the corresponding trace of Dirac matrices,
one finds
\bea
F_2(k^2)\big\vert_{\scriptsize{\mbox{1 loop}}}
 &=& ie^2\,\frac{32m_e^2}{k^2(k^2-4m_e^2)^2}\,\int\frac{d^4q}{(2\pi)^4}
\,\frac{1}{(p\,'+q)^2-m_e^2}\,\frac{1}{(p+q)^2-m_e^2}\,\frac{1}{q^2}
\nonumber\\
&&\qquad\times
\left[  
-3k^2(p\cdot q)^2 \,+\, 2k^2m_e^2(p\cdot q) \,+\, k^2m_e^2q^2 \,-\,m_e^2(k\cdot q)^2
\right]
\,.
\eea
Then follow the usual steps of introducing two Feynman parameters,
of performing a trivial change of variables and a symmetric integration 
over the loop momentum $q$, so that one arrives at
\bea
F_2(k^2) \big\vert_{\scriptsize{\mbox{1 loop}}}
&=& ie^2\,\frac{64m_e^2}{(k^2-4m_e^2)^2}\,
\int_0^1 dx x\int_0^1 dy \int\frac{d^4q}{(2\pi)^4}
\frac{1}{(q^2 - {\cal R}^2)^3}
\nonumber\\
&&\qquad\times
\left[
2x(1-x)m_e^4 \,-\, \frac{3}{4}x^2y^2(k^2)^2 \,+\,
m_e^2k^2x\left(3xy - y +\frac{1}{2}x\right)\right]
\nonumber\\
&=& 
\frac{e^2}{\pi^2}\,\frac{2m_e^2}{(k^2-4m_e^2)^2}\,
\int_0^1 dx x\int_0^1 dy \,\frac{1}{{\cal R}^2}\,
\nonumber\\
&&\qquad\times
\left[
2x(1-x)m_e^4 \,-\, \frac{3}{4}x^2y^2(k^2)^2 \,+\,
m_e^2k^2x\left(3xy - y +\frac{1}{2}x\right)\right]
\,,
\eea
with
\be
{\cal R}^2 \,=\, x^2y(1-y)(2m_e^2 - k^2) \,+\, x^2y^2m_e^2 \,+\, x^2(1-y)^2m_e^2
\,.
\ee
As expected, the limit $k^2\to 0$ can be taken without problem, and
gives
\be
a_e\vert_{\scriptsize{\mbox{1 loop}}}\equiv
F_2(0)\big\vert_{\scriptsize{\mbox{1 loop}}} \,=\,\frac{1}{2}\,\frac{\alpha}{\pi}
\,.
\lbl{schwinger1loop}
\ee
Let us stress that although the integral \rf{GammaFig1} diverges,
we have obtained a finite result for $F_2(k^2)$, and hence for $a_e$,  without
introducing any regularization. This is of course expected, since a divergence
in $F_2(0)$ would require that a counterterm of the form given by the 
second term in ${\widehat{\cal L}}_{\scriptsize{\mbox{int}}}$, see Eq.~\rf{Lintmod},
be introduced.
This would in turn spoil the renormalizability of the theory. In fact, as is 
well known, the divergence lies in $F_1(0)$, and goes into the renormalization
of the charge of the electron.

\indent

\begin{figure}[!h]
\centerline{\psfig{figure=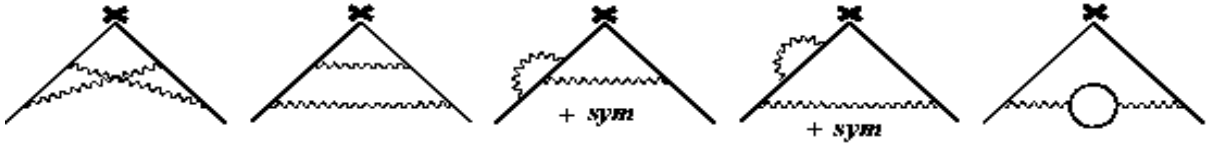,height=4cm,width=16cm}}
\caption{The Feynman diagrams which contribute to the coefficient
$A_2$ in Eq. {\protect\rf{a_lQED}}.} \label{fig:QED2loop}
\end{figure}

\indent


\subsection{Higher order mass independent corrections}

The previous calculation is rather straightforward and amounts to
the result
\be
A_1 \,=\, \frac{1}{2}
\ee
first obtained by Schwinger \cite{Schwinger48}.   Schwinger's
calculation was soon followed by a computation of $A_2$ \cite{KarplusKroll50}, 
which requires the evaluation of 7 graphs, representing five distinct
topologies, and shown in Fig.~\ref{fig:QED2loop}. Historically, the result 
of 
Ref. \cite{KarplusKroll50} was important, because it provided
the first explicit example of the realization of the renormalization
program of QED at two loops. However, the value for $A_2$ was not
given correctly.
The correct expression of
the second order mass independent contribution was derived in  
\cite{Peterman57a,Sommerfield57, Sommerfield58} (see also
\cite{Adkins89,SchwingerIII})
and reads \footnote{Actually, the experimental result of 
Ref. \cite{FrankenLiebes56} disagreed with the value
$A_2= -2.973$ obtained in \cite{KarplusKroll50}, and prompted
theoreticians to reconsider the calculation. The result
obtained by the authors of Refs. \cite{Peterman57a,Sommerfield57,Sommerfield58}
reconciled theory with experiment.}
\bea
A_2 &=&
\frac{197}{144} + \left( \frac{1}{2} - 3\ln 2\right) \zeta(2) + \frac{3}{4} \zeta(3)
\nonumber\\
&=&
- 0.328\,478\,965...
\eea
with $\zeta(p)={\displaystyle{\sum_{n=1}^\infty 1/n^p}}$, $\zeta(2)=\pi^2/6$.
The occurence of transcendental numbers like $\zeta(p)$ is a general
feature of higher order calculations in perturbative quantum field theory.
The pattern of these transcendentals
in perturbation theory and the structure of the renormalization
algorithm have also been put in relationship with other
mathematical structures, like knot theory and braids \cite{knot}, 
Hopf algebras \cite{hopf} and non commutative geometry \cite{gnc}.

\indent

\begin{figure}[!h]
\centerline{\psfig{figure=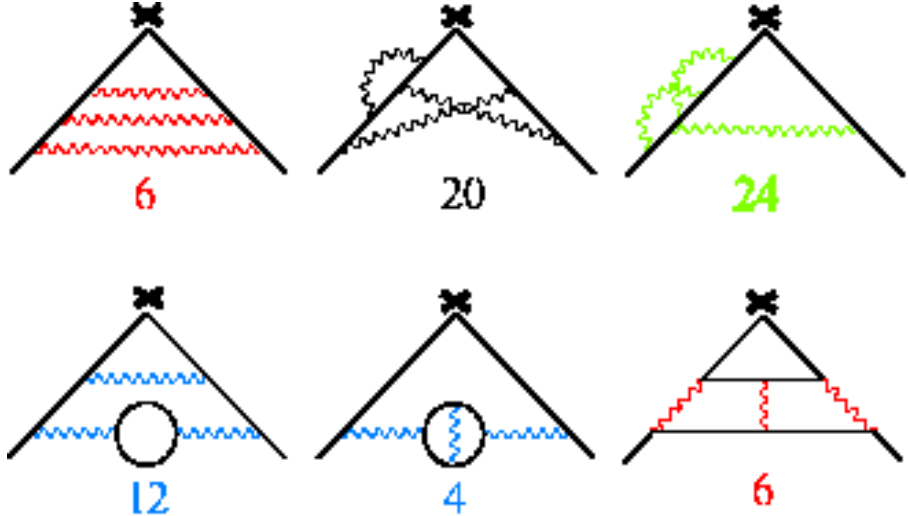,height=7cm,width=12cm}}
\caption{The 72 Feynman diagrams which make up the coefficient
$A_3$ in Eq. {\protect\rf{a_lQED}}.} \label{fig:QED3loop}
\end{figure}

\noindent
The analytic evaluation of the three loop mass independent contribution 
to the anomalous magnetic moment required
quite some time, and is mainly due to the dedication of E. Remiddi and his coworkers
during the period 1969-1996.
There are now 72 diagrams to consider, involving many different topologies, 
see Fig.~\ref{fig:QED3loop}.

\indent

\indent

\noindent
The calculation was completed \cite{LaportaRemiddi96} in 1996, 
with the analytical evaluation of a last class of diagrams,
the non planar ``triple cross" topologies. The result reads
\footnote{The completion of this three-loop program can be followed 
through 
Refs. 
\cite{BarbieriRemiddi75}-\cite{LaportaRemiddi95} and \cite{LaportaRemiddi96}.
A description of the technical aspects related to this
work and an account of its status up to 1990, with references to the
corresponding literature, are given in Ref. \cite{RoskiesRemiddiLevineQED}.}
\bea
A_3 &=&
\frac{83}{72}\pi^2\zeta(3) - \frac{215}{24}\zeta(5) + 
\frac{100}{3}\left[ \left(a_4 + \frac{1}{24}\ln^4 2\right) - \frac{1}{24}\pi^2\ln^2 2\right]
\nonumber\\
&&
\,- \frac{239}{ 2160}\pi^4 + \frac{139}{18}\zeta(3) - \frac{298}{9}\pi^2\ln 2 + \frac{17101}{ 810}\pi^2
+ \frac{28259}{ 5184}
\nonumber\\
&=&
1.181\,241\,456...
\eea
where~\footnote{The first three values are known to 
be $a_1=\ln2$, $a_2={\mbox{Li}}_2(1/2)=(\zeta(2)-\ln^2 2)/2$, $a_3=\frac{7}{8}\zeta(3) - \frac{1}{2}\zeta(2)\ln 2 +
\frac{1}{6}\ln^3 2$ \cite{RoskiesRemiddiLevineQED}.} $a_p={\displaystyle{\sum_{n=1}^\infty \frac{1}{2^n n^p}}}$.
The numerical value extracted from the exact analytical expression given above can be improved
to any desired order of precision. 

\indent

\noindent
In parallel to these analytical calculations, numerical methods for the 
evaluation of the higher order contributions were also developed, 
in particular by 
Kinoshita and his collaborators (for details, see \cite{KinoshitaQED}). 
The numerical evaluation of the full set of three loop diagrams was
achieved in several steps \cite{Aldinsetal70}-\cite{CvitanovicKinoshita74}.
The value quoted in \cite{CvitanovicKinoshita74} is $A_3=1.195(26)$, 
where the error comes from the numerical procedure. 
In comparison, let us quote the value \cite{KinoshitaLindquist, KinoshitaQED} 
$A_3=1.176\,11\,(42)$ obtained  if only a subset of 21 three loop
diagrams out of the original set of 72 is evaluated numerically, relying on the
analytical results for the remaining 51 ones, and recall the value 
$A_3=1.181\,241\,456...$ obtained from the full analytical evaluation.
The error induced on $a_e$ due to the numerical uncertainty in the second,
more accurate, value is still 
$\Delta(a_e) = 5.3\times 10^{-12}$, whereas  the experimental error is only
$\Delta(a_e)\vert_{\scriptsize{\mbox{exp}}} = 4.3\times 10^{-12}$. 
This discussion shows that the analytical evaluations of higher loop contributions
to the anomalous magnetic moment of the electron have a strong practical interest as
far as the precision of the theoretical prediction is concerned, and which goes well
beyond the mere intellectual satisfaction and technical skills involved in these
calculations.
\footnote{It is only fair to point out that the numerical
values that are quoted here correspond to those given in 
the original references. It is to be expected that they would 
certainly improve if today's numerical possibilities were used.}

\indent

\noindent
At the four loop level, there are 891 diagrams to consider. 
Clearly, only a few of them
have been evaluated analytically \cite{Caffoetal84,RemiddiSorella85}.
The complete numerical evaluation of the whole set gave \cite{KinoshitaLindquist}
$A_4= -1.434(138)$. The developement of computers allowed
subsequent reanalyses to be more accurate, i.e.
$A_4= -1.557(70)$ \cite{Kinoshita95}. Until recently, 
the ``latest of [these] 
constantly improving values" was \cite{HughesKinoshita99}
$A_4 \,=\,-1.509\,8(38\,4)$.
This calculation certainly represents a formidable task, and
requires many elaborate technical tools. A descriptive 
account can be found in \cite{KinoshitaQED}. Let us mention, 
for completeness, that efforts to improve upon the evaluation of 
$A_4$ are presently being pursued. Thus, a mistake has recently been found
in an earlier computer code used for the evaluation of a subclass
of four loop diagrams \cite{KinoshitaNio03}, whose contribution
to $A_4$ was $A_{4;IV(d)}= -0.7503(60)$ [for the precise meaning 
of the notation, we refer the reader to Refs. \cite{KinoshitaQED} 
and \cite{KinoshitaLindquist}]. The corrected value reads instead 
\cite{KinoshitaNio03} $A_{4;IV(d)} = -0.99072(10)$ [note also the 
impressive improvement in the precision]. By itself, this correction
induces a 16\% downward shift of the value of $A_4$, 
a far from trivial modification [see below]. At the present stage,
the value of $A_4$ reads
\be
A_4 \,=\,-1.750\,2(38\,4)
\,.
\ee

\indent

\noindent
Needless to say, so far the 
five loop contribution $A_5$ is unknown territory. 
On the other hand, $(\alpha/\pi)^5\sim 7\times 10^{-14}$,
so that one may reasonably expect, in view of the present 
experimental situation,  that its knowledge is not yet required.

\subsection{Mass dependent QED corrections}

\begin{figure}[!h]
\centerline{\psfig{figure=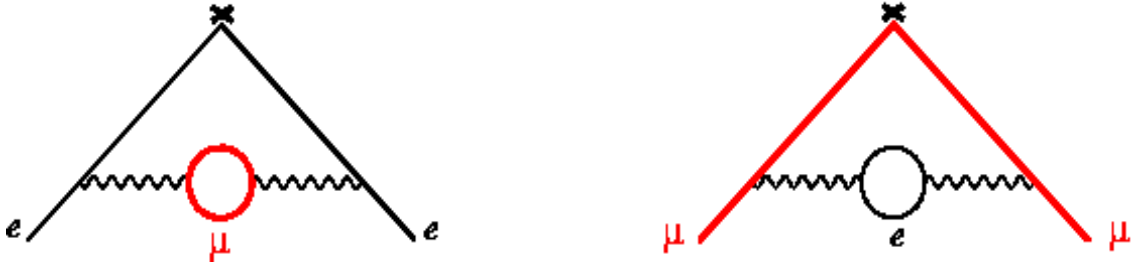,height=3.5cm,width=15cm}}
\caption{The insertion of a muon vacuum polarization loop into
the electron vertex correction (left) or of an electron vacuum polarization 
loop into the muon vertex correction (right).} \label{fig:vacpol}
\end{figure}

We now turn to the QED loop contributions to the electron's 
anomalous magnetic moment involving the heavier
leptons, $\mu$ and $\tau$. The lowest order contribution 
of this type occurs at the two loop level, ${\cal O}(\alpha^2)$, and 
corresponds to a heavy lepton vacuum polarization insertion 
in the one loop vertex graph, cf. Fig. \ref{fig:vacpol}.
Quite generally, the contribution to $a_\ell$ arising from
the insertion, into the one loop vertex correction, of 
a vacuum polarization graph due to a loop of lepton $\ell\,'$,
reads \cite{SuuraWichman57,Peterman57b} \footnote{A trivial
change of variable on $s$ brings the expression \rf{B2disp}
into the form given in \cite{SuuraWichman57,Peterman57b}.
Furthermore, the analytical result obtained upon
performing the double integration is available in \cite{Elend66}.}
\be
B_2(\ell,\ell\,' ) \,=\, \frac{1}{3}\,\int_{4m_{\ell\,'}^2}^{\infty}ds
\sqrt{1 - \frac{4m_{\ell\,'}^2}{s}}\,\frac{s+2m_{\ell\,'}^2}{s^2}\,
\int_0^1 dx\,\frac{x^2(1-x)}{x^2 + (1-x)\,\frac{s}{m_{\ell}^2}}
\,.
\lbl{B2disp}
\ee

\noindent
If $m_{\ell\,'}\gg m_{\ell}$, the second integrand can be approximated by
$x^2m_{\ell}^2/s$, and one obtains \cite{LautrupdeRafael68}
\be
B_2(\ell,\ell\,' ) \,=\, \frac{1}{45}\,\left(\frac{m_\ell}{m_{\ell\,'}}\right)^2 \,+\,\cdots 
\,,\ m_{\ell\,'} \gg m_\ell\,.
\ee
The complete expansion of $B_2(\ell,\ell\,' )$ for 
$m_{\ell\,'} \gg m_\ell$ can be found in \cite{SamuelLi91},
from which we quote
\bea
B_2(\ell,\ell\,' ) &=& \frac{1}{45}\,\left(\frac{m_\ell}{m_{\ell\,'}}\right)^2 \,+\,
\frac{1}{70}\,\left(\frac{m_\ell}{m_{\ell\,'}}\right)^4\,
\ln\left(\frac{m_\ell}{m_{\ell\,'}}\right)
\,+\, 
\frac{9}{19600}\,\left(\frac{m_\ell}{m_{\ell\,'}}\right)^4 
\nonumber\\
&&
+\,
\frac{4}{315}\,\left(\frac{m_\ell}{m_{\ell\,'}}\right)^6\,
\ln\left(\frac{m_\ell}{m_{\ell\,'}}\right)
\,-\,
\frac{131}{99225}\,\left(\frac{m_\ell}{m_{\ell\,'}}\right)^6
\nonumber\\
&&
+\,
{\cal O}\left[ \left(\frac{m_\ell}{m_{\ell\,'}}\right)^8
\ln\left(\frac{m_\ell}{m_{\ell\,'}}\right)\right]
\,.
\lbl{B2dispexp}
\eea

\indent

\indent

\noindent
Numerically, this translates into 
[the values for the masses that were used read
$m_e=0.510\,998\,902(21)$ MeV, 
$m_\tau = 1\,776.99^{+0.29}_{-0.26}$ \cite{PDG02}, and
$m_\mu/m_e = 206.768\,277(24)$ \cite{Liuetal99}]
\bea
B_2(e,\mu ) &=& 5.197 \times 10^{-7}
\nonumber\\
B_2(e,\tau ) &=& 1.838 \times 10^{-9}
\,.
\eea

\indent

\noindent
For later use, it is interesting to briefly discuss the structure of Eq. \rf{B2disp}. The quantity
which appears under the integral is related to the cross section for the
scattering of a $\ell^+\ell^-$ pair into a pair $(\ell\,')^+(\ell\,')^-$ {\em at lowest
order in QED},
\be
\sigma_{\scriptsize{\mbox QED}}^{\ell^+\ell^-\to(\ell\,')^+(\ell\,')^-}(s) 
\,=\, \frac{4\pi \alpha^2}{3s^2}\,\sqrt{1 - \frac{4m_{\ell\,'}^2}{s}}\,(s+2m_{\ell\,'}^2)
\,,
\ee
so that
\be
B_2(\ell;\ell\,') \,=\,\frac{1}{3}\int_{4m_{\ell\,'}^2}^{\infty}
\frac{ds}{s}\,
K(s)R^{(\ell\,')}(s)
\,,
\ee
where \footnote{Explicit expressions for $K(s)$ are also available, but
for many purposes, the integral representation given here turns out
to be more convenient.}
\be
K(s) \,=\,\int_0^1 dx\,\frac{x^2(1-x)}{x^2 + (1-x)\,\frac{s}{m_{\ell}^2}}
\lbl{defK}
\,,
\ee
and $R^{(\ell\,')}(s)$ is the {\em lowest order} QED cross section 
$\sigma_{\scriptsize{\mbox QED}}^{\ell^+\ell^-\to(\ell\,')^+(\ell\,')^-}(s)$
divided by the asymptotic form of the cross section of the reaction $e^+e^-\to\mu^+\mu^-$
for $s\gg m_\mu^2$, $\sigma_\infty^{e^+e^-\to\mu^+\mu^-}(s) = \frac{4\pi \alpha^2}{3s}$.

\indent

\begin{figure}[!h]
\centerline{\psfig{figure=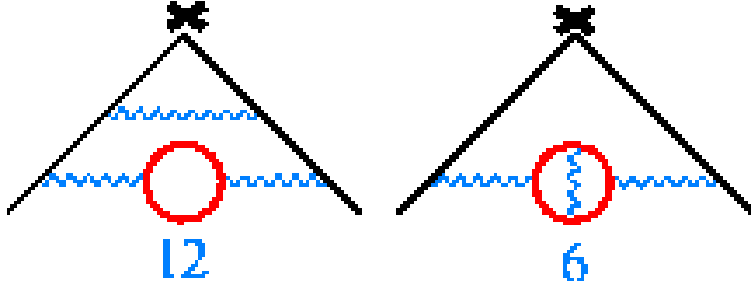,height=4cm,width=10cm}}
\caption{Three loop QED corrections with insertion of a heavy lepton 
vacuum polarization which make up the coefficient 
$B_3^{({\mbox{\scriptsize{v.p.}}})}(e;\mu)$.
} \label{fig:Sgmemup}
\end{figure}

\begin{figure}[!h]
\centerline{\psfig{figure=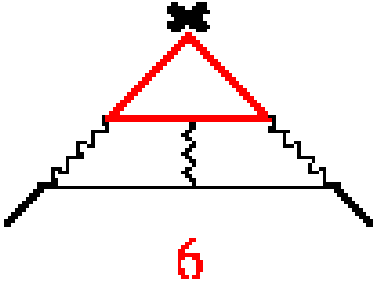,height=4cm,width=5cm}}
\caption{The three loop QED correction with the insertion of a heavy lepton 
light-by-light scattering subgraph, corresponding to the coefficient 
$B_3^{({\mbox{\scriptsize{L$\times$L}}})}(e;\mu)$.
} \label{fig:Sgmemul}
\end{figure}

\noindent
The three loop contributions with different lepton flavours in the
loops are also known analytically \cite{Laporta93b,LaportaRemiddi93}. 
It is convenient to distinguish
three classes of diagrams. The first group contains all the diagrams
with one or two vacuum polarization insertion involving the same
lepton, $\mu$ ot $\tau$, of the type shown in Fig. \ref{fig:Sgmemup}. 
The second group consists of the leptonic 
light-by-light scattering
insertion diagrams, Fig. \ref{fig:Sgmemul}. 
Finally, since there are three flavours of massive
leptons in the standard model, one has also the possibility of having 
graphs with two heavy lepton vacuum polarization insertions, one made 
of a muon loop, the other of a $\tau$ loop. This gives
\be
B_3(e,\ell) \,=\,B_3^{({\scriptsize{\mbox{v.p.}}})}(e;\mu) \,+\, 
B_3^{({\scriptsize{\mbox{v.p.}}})}(e;\tau)
\,+\,B_3^{({\scriptsize{\mbox{L$\times$L}}})}(e;\mu) \,+\, 
B_3^{({\scriptsize{\mbox{L$\times$L}}})}(e;\tau)
\,+\,B_3^{({\scriptsize{\mbox{v.p.}}})}(e;\mu, \tau)\,.
\ee
The analytical expression for $B_3^{({\scriptsize{\mbox{v.p.}}})}(e;\mu)$ 
can be found in Ref. \cite{Laporta93b}, whereas \cite{LaportaRemiddi93}
gives the corresponding result for
$B_3^{({\mbox{\scriptsize{L$\times$L}}})}(e;\mu)$. 
For practical purposes, it is both sufficient
and more convenient to use their expansions in powers of ${m_e}/{m_\mu}$,
\bea
B_3^{({\scriptsize{\mbox{v.p.}}})}(e;\mu) &=&
\left(\frac{m_e}{m_\mu}\right)^2
\left[-\frac{23}{135}\ln\left(\frac{m_\mu}{m_e}\right) - \frac{2}{45}\pi^2 + 
\frac{10117}{24300}\right]
\nonumber\\
&+&
\left(\frac{m_e}{m_\mu}\right)^4\left[\frac{19}{2520}\ln^2\left(\frac{m_\mu}{m_e}\right) - 
\frac{14233}{132300}\ln\left(\frac{m_\mu}{m_e}\right) + \frac{49}{768}\zeta(3) - \frac{11}{945}\pi^2 + \frac{2976691}{296352000}\right] 
\nonumber\\
&+& 
{\cal O}\left[ \left(\frac{m_e}{m_\mu}\right)^6\right]
\nonumber\\
&=& -0.000\,021\,768...
\eea

\indent

\indent

\noindent
and 
\bea
B_3^{({\mbox{\scriptsize{L$\times$L}}})}(e;\mu) &=&
\left(\frac{m_e}{m_\mu}\right)^2
\left[\frac{3}{2}\zeta(3)- \frac{19}{16}\right]
\nonumber\\
&+&
\left(\frac{m_e}{m_\mu}\right)^4\left[- \frac{161}{810}\ln^2\left(\frac{m_\mu}{m_e}\right) - 
\frac{16189}{48600}\ln\left(\frac{m_\mu}{m_e}\right) + \frac{13}{18}\zeta(3) - \frac{161}{9720}\pi^2 -\frac{831931}{972000}\right] 
\nonumber\\
&+&
{\cal O}\left[ \left(\frac{m_e}{m_\mu}\right)^6\right]
\nonumber\\
&=& 0.000\,014\,394\,5...
\eea

\indent

\indent

\noindent
The expressions for $B_3^{({\mbox{\scriptsize{v.p.}}})}(e;\tau)$ and
$B_3^{({\mbox{\scriptsize{L$\times$L}}})}(e;\tau)$ follow upon replacing
the muon mass $m_\mu$ by $m_\tau$. This again gives a suppression
factor $(m_\mu/m_\tau)^2$, which makes these contributions negligible
at the present level of precision. For the same reason, 
$B_3^{({\mbox{\scriptsize{v.p.}}})}(e;\mu, \tau)$ can also be
discarded.

\subsection{Other contributions to $a_e$}

In order to make the discussion of the standard model contributions
to $a_e$ complete, there remains to mention the hadronic and weak
components, $a_{e}^{\mbox{\scriptsize{had}}}$ and
$a_{e}^{\mbox{\scriptsize{weak}}}$, respectively. 
Their features will be discussed in detail below,
in the context of the anomalous magnetic moment of the muon.
I therefore only quote the numerical values \footnote{I
reproduce here the values given in \cite{CzarneckiMarciano98,HughesKinoshita99},
except for the fact that I have taken into account the changes
in the value of the hadronic light-by-light contribution to $a_\mu$,
see below, for which I take 
$a_{\mu}^{({\mbox{\tiny{L$\times$L}}})}=+8(4)\times 10^{-10}$,
and which translates into 
$a_e^{({\mbox{\tiny{L$\times$L}}})}\sim
a_\mu^{({\mbox{\tiny{L$\times$L}}})}(m_e/m_\mu)^2=0.02\times 10^{-12}$.}
\be
a_{e}^{\mbox{\scriptsize{had}}} \,=\, 1.67(3)\times 10^{-12}
\,,
\ee
and \cite{Czarneckietal96}
\be
a_{e}^{\mbox{\scriptsize{weak}}} \,=\, 0.030\times 10^{-12}
\,.
\ee

\indent

\subsection{Comparison with experiment and determination of $\alpha$}

Summing up the various contributions discussed so far gives
the standard model prediction 
\cite{CzarneckiMarciano98,HughesKinoshita99,Nyffeler03,KinoshitaNio03}
\be
a_e^{\mbox{\scriptsize{SM}}}\,=\,
0.5\,\frac{\alpha}{\pi}\,-\,0.328\,478\,444\,00\left(\frac{\alpha}{\pi}\right)^2
\,+\,1.181\,234\,017\left(\frac{\alpha}{\pi}\right)^3\,-\,
1.750\,2(38\,4)\left(\frac{\alpha}{\pi}\right)^4
+1.70(3)\times 10^{-12}
\,.
\ee
In order to obtain a number that can be compared to the experimental
result, a sufficiently accurate determination of the fine structure
constant $\alpha$ is required. The best available measurement of the 
latter comes from the quantum Hall effect \cite{MohrTaylorRMP00},
\be
\alpha^{-1}(qH)\,=\,137.036\,003\,00(2\,70)
\lbl{alphaqH}
\ee
and leads to
\be
a_e^{\mbox{\scriptsize{SM}}}(qH)\,=\,0.001\,159\,652\,146\,5(24\,0)
\,,
\ee
about six times less accurate than the experimental value 
of Eq. \rf{a_eexpaverage}, 
$a_e^{\mbox{\scriptsize{exp}}}\,=\,0.001\,159\,652\,188\,3(4\,2)$.
On the other hand, if one excludes other contributions to $a_e$
than those from the standard model considered so far, i.e. 
if one identifies $a_e^{\mbox{\scriptsize{SM}}}$ with
$a_e^{\mbox{\scriptsize{exp}}}$, 
then the value of $a_e^{\mbox{\scriptsize{exp}}}$  
as given in Eq. \rf{a_eexpaverage} provides
the best determination of $\alpha$ to date 
\cite{KinoshitaNio03,Nyffeler03},
\be
\alpha^{-1}(a_e)\,=\,137.035\,998\,75(52)
\,,
\lbl{alphafroma_e}
\ee
at least to the extend that one may reasonably
believe that all theoretical errors are under control.
Now, as we have seen earlier, the value of $A_4$ has
recently been corrected \cite{KinoshitaNio03}. The analysis
presented here incorporates the changes brought forward
by the analysis of Ref. \cite{KinoshitaNio03}. It lowers the
prediction for $a_e$ by $\sim 7.0\times 10^{-12}$ if
one uses the value \rf{alphaqH} of $\alpha$, or equivalently
reduces the value of $\alpha^{-1}(a_e)$ by one and a half
standard deviation. It is very likely that the completion
of the analysis begun in Ref. \cite{KinoshitaNio03} will
lead to a more accurate determination of $A_4$, and further
changes in the numbers quoted here are to be expected.

\indent

\begin{center}
{{
Exercises for section 4
}}
\end{center}

\noindent
{\it Exercise 4.1}\\
Reproduce the steps that lead from Eq. \rf{GammaFig1}
to Eq. \rf{schwinger1loop}.\\

\indent

\section{The anomalous magnetic moment of the muon}
\renewcommand{\theequation}{\arabic{section}.\arabic{equation}}
\setcounter{equation}{0}

In this section, the theoretical aspects concerning the 
anomalous magnetic moment of the muon are discussed. 
Since the muon is much heavier than the electron, $a_\mu$ will be more sensitive to higher
mass scales. In particular, it is a better probe for possible degrees of freedom
beyond the standard model, like supersymmetry. The drawback, however,
is that $a_\mu$ will also be more sensitive to the non perturbative strong
interaction dynamics at the $\sim 1$ GeV scale.

\subsection{QED contributions to $a_\mu$}

As already mentioned before, the mass independent QED contributions to $a_\mu$
are described by the same coefficients $A_n$ as in the case of the electron. We 
therefore need only to discuss the coefficients $B_n(\mu;\ell\,')$ associated
with the mass dependent corrections.

\indent

\noindent
For $m_{\ell\,'}\ll m_\ell$, Eq. \rf{B2disp} gives 
\cite{SuuraWichman57,Peterman57b,Elend66,SamuelLi91}
\be
B_2(\ell;\ell\,')\,=\,
\frac{1}{3}\,\ln\left(\frac{m_\ell}{m_{\ell '}}\right) \,-\, \frac{25}{36} \,+\,
\frac{\pi^2}{4}\,\frac{m_{\ell '}}{m_{\ell}}
\,-\,4 \left(\frac{m_{\ell '}}{m_{\ell}}\right)^2 \ln\left(\frac{m_\ell}{m_{\ell '}}\right) \, +\,
3\left(\frac{m_{\ell '}}{m_{\ell}}\right)^2 \,+\,
{\cal O}\left[ \left(\frac{m_{\ell '}}{m_{\ell }}\right)^3
 \right]
\,.
\ee
The complete expansion in powers of $m_{\ell\,'}/ m_\ell$ can
again be found in \cite{SamuelLi91}.
In the case of $B_2(\mu;\tau)$, one 
may use the expression given in
Eq. \rf{B2dispexp}.
Upon using the values $m_e = 0.510\,998\,902(21)$ MeV,
$m_\mu = 105.658\,357(5)$ MeV, $m_\tau = 1776.99^{+0.29}_{-0.26}$ MeV
 \cite{PDG02}, and $m_\mu / m_e = 206.768\,277(24)$ \cite{Liuetal99}
the corresponding numbers read 
\be
B_2(\mu;e) \,=\,1.094\,258\,300(38)
\ee
\be
B_2(\mu;\tau) \,=\,0.000\,078\,064(25)
\,.
\ee
Although these numbers follow from an 
analytical expression, there are uncertainties attached to them,
induced by those on the corresponding values 
of the ratios of the lepton masses.

\indent

\noindent
The three loop QED corrections decompose as 
\be
B_3(\mu,\ell) \,=\,B_3^{({\mbox{\scriptsize{v.p.}}})}(\mu;e) \,+\, 
B_3^{({\mbox{\scriptsize{v.p.}}})}(\mu;\tau)
\,+\,B_3^{({\mbox{\scriptsize{L$\times$L}}})}(\mu;e) \,+\, 
B_3^{({\mbox{\scriptsize{L$\times$L}}})}(\mu;\tau)
\,+\,B_3^{({\mbox{\scriptsize{v.p.}}})}(\mu;e, \tau)\,,
\ee
with \cite{Laporta93b,LaportaRemiddi93}
\bea
&&
B_3^{({\mbox{\scriptsize{v.p.}}})}(\mu;e) \,=\,
\frac{2}{9}\,\ln^2\left(\frac{m_\mu}{m_e}\right)\,+\,
\left[
\zeta(3) - \frac{2}{3}\pi^2\ln 2 + \frac{1}{9}\pi^2 + \frac{31}{27}
\right]
\ln\left(\frac{m_\mu}{m_e}\right)
\nonumber\\
&&
+ \,\frac{11}{216}\pi^4 \,-\, \frac{2}{9}\pi^2\ln^2 2 \,-\,\frac{8}{3}a_4
\,-\, \frac{1}{9}\ln^4 2 \,-\,3\zeta(3) \,+\, \frac{5}{3}\pi^2\ln 2 
\,-\, \frac{25}{18}\pi^2 \,+\, \frac{1075}{216}
\nonumber\\
&&
+\,\frac{m_e}{m_\mu}\,
\left[
-\frac{13}{18}\pi^3 - \frac{16}{9}\pi^2\ln 2 + \frac{3199}{1080}\pi^2
\right]
\nonumber\\
&&
+\,\left(\frac{m_e}{m_\mu}\right)^2\,
\left[
\frac{10}{3}\ln^2\left(\frac{m_\mu}{m_e}\right) - \frac{11}{9}\ln\left(\frac{m_\mu}{m_e}\right)
- \frac{14}{3}\pi^2\ln 2 - 2\zeta(3)+ \frac{49}{12}\pi^2 - \frac{131}{54}
\right]
\nonumber\\
&&
+\,\left(\frac{m_e}{m_\mu}\right)^3\,
\left[
\frac{4}{3}\pi^2\ln\left(\frac{m_\mu}{m_e}\right) + \frac{35}{12}\pi^3 - \frac{16}{3}\pi^2\ln 2 - 
\frac{5771}{1080}\pi^2
\right]
\nonumber\\
&&
+\,\left(\frac{m_e}{m_\mu}\right)^4\,
\bigg[
- \frac{25}{9}\ln^3\left(\frac{m_\mu}{m_e}\right) - \frac{1369}{180}\ln^2\left(\frac{m_\mu}{m_e}\right)
\nonumber\\
&&
+ [ -2\zeta(3) + 4\pi^2\ln 2 - \frac{269}{144}\pi^2 - \frac{7496}{675}] \ln\left(\frac{m_\mu}{m_e}\right)
\nonumber\\
&&
- \,\frac{43}{108}\pi^4 \,+\, \frac{8}{9}\pi^2\ln^2 2 \,+\,\frac{80}{3}a_4
\,+\, \frac{10}{9}\ln^4 2 \,-\,\frac{411}{32}\zeta(3) \,+\, \frac{89}{48}\pi^2\ln 2 
\,-\, \frac{1061}{864}\pi^2 \,-\, \frac{274511}{54000}
\bigg]
\nonumber\\
&&
+ \, {\cal O}\bigg[\left(\frac{m_e}{m_\mu}\right)^5\bigg]
\,,
\eea
\bea
&&
B_3^{({\mbox{\scriptsize{L$\times$L}}})}(\mu;e) \,=\,
\frac{2}{3}\,\pi^2\,\ln\left(\frac{m_\mu}{m_e}\right)\,+\,\frac{59}{270}\,\pi^4 \,-\,3\zeta(3)
\,-\,\frac{10}{3}\,\pi^2 \,+\,\frac{2}{3}
\nonumber\\
&&
+\,\frac{m_e}{m_\mu}\,
\left[
\frac{4}{3}\pi^2\ln\left(\frac{m_\mu}{m_e}\right)  - \frac{196}{3}\pi^2\ln 2 + 
\frac{424}{9}\pi^2
\right]
\nonumber\\
&&
+\,\left(\frac{m_e}{m_\mu}\right)^2\,
\bigg[
- \frac{2}{3}\ln^3\left(\frac{m_\mu}{m_e}\right)  +
(\frac{\pi^2}{9} - \frac{20}{3})\ln^2\left(\frac{m_\mu}{m_e}\right) -
[ \frac{16}{135}\pi^4 +4\zeta(3) - \frac{32}{9}\pi^2 + \frac{61}{3} ] \ln\left(\frac{m_\mu}{m_e}\right) 
\nonumber\\
&&
+ \frac{4}{3}\zeta(3)\pi^2 - \frac{61}{270}\pi^4 + 3\zeta(3) +  \frac{25}{18}\pi^2 - \frac{283}{12}
\bigg]
\nonumber\\
&&
+\,\left(\frac{m_e}{m_\mu}\right)^3\,
\left[
\frac{10}{9}\pi^2\ln\left(\frac{m_\mu}{m_e}\right) \,-\, \frac{11}{9}\pi^2
\right]
\nonumber\\
&&
+\,\left(\frac{m_e}{m_\mu}\right)^4\,
\bigg[
 \frac{7}{9}\ln^3\left(\frac{m_\mu}{m_e}\right) + \frac{41}{18}\ln^2\left(\frac{m_\mu}{m_e}\right)
+ \frac{13}{9}\pi^2\ln\left(\frac{m_\mu}{m_e}\right) + \frac{517}{108}\ln\left(\frac{m_\mu}{m_e}\right)
\nonumber\\
&&
+ \frac{1}{2}\zeta(3)  + \frac{191}{216}\pi^2 + 
\frac{13283}{2592}
\bigg]
\, + \, {\cal O}\bigg[\left(\frac{m_e}{m_\mu}\right)^5\bigg]
\,,
\eea
while $B_3^{({\mbox{\scriptsize{v.p.}}})}(\mu ;\tau)$ and 
$B_3^{({\mbox{\scriptsize{L$\times$L}}})}(\mu ;\tau)$ are
derived from $B_3^{({\mbox{\scriptsize{v.p.}}})}(e ;\mu)$ and from
$B_3^{({\mbox{\scriptsize{L$\times$L}}})}(e; \mu)$, respectively,
by trivial substitutions of the masses. 
Furthermore, the graphs with mixed vacuum polarization
insertions, one electron loop, and one $\tau$ loop, are evaluated
numerically using a dispersive integral 
\cite{SamuelLi91,Laporta93b,Krause97}.

\indent

\noindent
Numerically, one obtains 
\bea
B_3^{({\mbox{\scriptsize{v.p.}}})}(\mu ;e) &=& 1.920\,455\,1(2)
\nonumber\\
B_3^{({\mbox{\scriptsize{L$\times$L}}})}(\mu ;e) &=& 20.947\,924\,7(7)
\nonumber\\
B_3^{({\mbox{\scriptsize{v.p.}}})}(\mu ;\tau) &=& -0.001\,782\,3(5)
\nonumber\\
B_3^{({\mbox{\scriptsize{L$\times$L}}})}(\mu ;\tau) &=& 0.002\,142\,9(7)
\nonumber\\
B_3^{({\mbox{\scriptsize{v.p.}}})}(\mu ;e, \tau) &=& 0.000\,527\,7(2)
\,.
\eea
Notice the large value of 
$B_3^{({\mbox{\scriptsize{L$\times$L}}})}(\mu ;e)$,
due to the occurence of terms involving factors like
$\ln(m_\mu/m_e)\sim 5$ and  powers of $\pi$. Such a large contribution,
first obtained numerically in Ref. \cite{Aldinsetal70}, allowed to
explain a discrepancy of 1.7 $\sigma$ between the theoretical value 
and the experimental measurement of Ref. \cite{Baileyetal68}. Finally,
several
pieces of $B_3^{({\mbox{\scriptsize{v.p.}}})}(\mu ;e)$ had already
been worked out earlier, in Refs. 
\cite{Kinoshita67,LautrupdeRafael68,Lautrup70,LautrupdeRafael70,LautrupPetermandeRafael71}

\indent

\noindent
The contributions at fourth order in $\alpha$ have been obtained
numerically in Ref. \cite{Kinoshita93}. They must be corrected
for the change in $A_4$ obtained in \cite{KinoshitaNio03}.
At the next order, no full calculation, even through numerical 
techniques, is available. Specific contributions, for instance those
enhanced by powers of $\ln (m_e/m_\mu)$ times powers
of $\pi$, have been evaluated 
\cite{Kinoshitaetal90,Yelkhovsky89,MilsteinYelkhovsky89,Karshenboim93}.

\indent

\noindent
Putting all these contributions together leads to
the expression
\be
a_\mu^{\mbox{\scriptsize{QED}}}\,=\,
0.5\,\frac{\alpha}{\pi}\,+\,
0.765\,857\,399(45)\left(\frac{\alpha}{\pi}\right)^2
\,+\,24.050\,509\,5(2\,3)\left(\frac{\alpha}{\pi}\right)^3
\,+\, 125.08(41)\left(\frac{\alpha}{\pi}\right)^4
\,+\, 930(170)\left(\frac{\alpha}{\pi}\right)^5
\,.
\ee
Upon inserting the value of $\alpha$ obtained from
the anomalous magnetic moment of the electron in
Eq. \rf{alphafroma_e}, one finds
\be
a_\mu^{\mbox{\scriptsize{QED}}}\,=\,11\, 658\, 470.35(28)
\times 10^{-10}
\,.
\ee

\subsection{Hadronic contributions to $a_\mu$}

On the level of Feynman diagrams, hadronic contributions arise
through loops of virtual quarks and gluons. These loops also 
involve the soft scales, and therefore cannot be computed
reliably in perturbative QCD. 
We shall decompose the hadronic contributions into three subsets:
hadronic vacuum polarization insertions at order $\alpha^2$, at 
order $\alpha^3$, and hadronic light-by-light scattering,
\be
a_{\mu}^{\mbox{\scriptsize{had}}} \,=\, a_{\mu}^{({\mbox{\scriptsize{h.v.p. 1}}})} \,+\,
a_{\mu}^{({\mbox{\scriptsize{h.v.p. 2}}})}\,+\,
a_{\mu}^{({\mbox{\scriptsize{h. L$\times$L}}})}
\ee

\subsubsection{Hadronic vacuum polarization}

\begin{figure}[!h]
\centerline{\psfig{figure=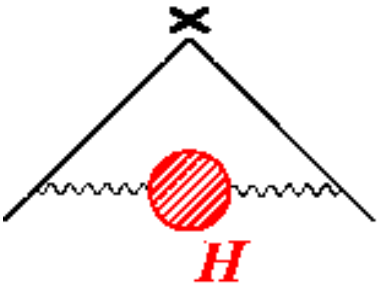,height=4cm,width=5cm}}
\caption{The insertion of the hadronic vacuum 
polarization into the one loop vertex
correction, corresponding to 
$a_{\mu}^{({\mbox{\scriptsize{h.v.p. 1}}})}$.
} \label{fig:amuhadrp}
\end{figure}

We first discuss $a_{\mu}^{({\mbox{\scriptsize{h.v.p. 1}}})}$, 
which arises at order ${\cal O}(\alpha^2)$ from the insertion of a single hadronic vacuum
polarization into the lowest order vertex correction graph, see
Fig.~\ref{fig:amuhadrp}. The importance
of this contribution to $a_\mu$ is known since long time 
\cite{BouchiatMichel61,DurandIII62}.

\indent

\indent

\noindent
There is a 
very convenient dispersive representation of this diagram,
similar to Eq. \rf{B2disp}
\bea
a_{\mu}^{({\mbox{\scriptsize{h.v.p. 1}}})} &=& 4{\alpha}^2\,
\int_{4M_\pi^2}^{\infty} \,\frac{ds}{s}\,K(s)\,\frac{1}{\pi}\,{\mbox{Im}}\Pi(s)
\nonumber\\
&=&
\frac{1}{3}\,\left(\frac{\alpha}{\pi}\right)^2\,
\int_{4M_\pi^2}^{\infty} \,\frac{ds}{s}\,K(s)R^{\mbox{\scriptsize{had}}}(s)
\,,
\lbl{hvp1}
\eea
Here, $\Pi(s)$ denotes the {\em hadronic} vacuum 
polarization
function, defined as \footnote{Actually, $\Pi(s)$ defined this
way has an ultraviolet divergence, produced by the QCD short distance
singularity of the chronological product of the two currents.
However, it only affects the real part of $\Pi(s)$. 
A renormalized, finite quantity is obtained
by a single subtraction, $\Pi(s) - \Pi(0)$.}
\be
(q_\mu q_\nu - q^2\eta_{\mu\nu}) \Pi(Q^2)
\,=\,i
\int d^4x e^{iq\cdot x}\,\langle\Omega\vert \mbox{T}\{
j_{\mu}(x)j_\nu(0)\}\vert\Omega\rangle
\,,
\ee
with $j_{\rho}$ the hadronic component of the electromagnetic current, 
$Q^2=-q^2\ge 0$ for $q_\mu$ spacelike, and
$\vert\Omega\rangle$ the QCD vacuum. The function
$K(s)$ was defined in Eq. \rf{defK}, and 
$R^{\mbox{\scriptsize{had}}}(s)$ stands now [see however below]
 for the cross 
section $\sigma_0^{e^+e^-\to\mbox{\scriptsize{had}}}(s)$ of 
$e^+e^-\to \mbox{hadrons}$, {\em at lowest order in $\alpha$}, divided by 
$\sigma_\infty^{e^+e^-\to\mu^+\mu^-}(s) = \frac{4\pi \alpha^2}{3s}$.
A first principle computation of this strong interaction contribution
is far beyond our present abilities to deal with the non perturbative
aspects of confining gauge theories. 
This last relation is however very interesting because it expresses 
$a_{\mu}^{({\mbox{\scriptsize{h.v.p. 1}}})}$
through a quantity that can be measured experimentally. In this respect,
two important properties of the function $K(s)$ deserve to be mentioned.
First, it appears from the integral representation \rf{defK} that $K(s)$
is positive definite. Since $R^{\mbox{\scriptsize{had}}}$ 
is also positive, one deduces that
$a_{\mu}^{({\mbox{\scriptsize{h.v.p. 1}}})}$ itself is positive. Second, the 
function $K(s)$ decreases as $m_\mu^2/3s$ as $s$ grows, so that it is indeed
the low energy region which dominates the integral. Explicit evaluation
of $a_{\mu}^{({\mbox{\scriptsize{h.v.p. 1}}})}$ using available data
actually reveals that more than 80\% of its value comes from energies
below 1.4 GeV. This observation is rather welcome, since the spectral
density ${\mbox{Im}}\Pi(s)$ can also be extracted from data on the
hadronic decays of the $\tau$ lepton in this energy region.

\indent

\noindent
Finally, the values obtained in this way for 
$a_{\mu}^{({\mbox{\scriptsize{h.v.p. 1}}})}$
have evolved in time, as shown in Table \ref{tab:tab3}. 
This evolution is mainly 
driven by the availability of more data, and is still going on, 
as the last entries
of Table \ref{tab:tab3} show. 
In order to match the precision reached by the latest experimental
measurement of $a_\mu$, $a_{\mu}^{({\mbox{\scriptsize{h.v.p. 1}}})}$
needs to be known at $\sim 1\%$.
Besides the very recent high quality
$e^+e^-$ data obtained by the BES Collaboration
\cite{BES}
in the region between 2 to 5 GeV, 
and by the CMD-2 collaboration \cite{CMD-2} in the 
region dominated by the $\rho$ resonance, 
the latest analyses sometimes also include or use, in the 
low-energy region, data obtained from hadronic decays of the 
$\tau$ by ALEPH \cite{ALEPH}, 
and, more recently, by OPAL \cite{OPAL99} and CLEO \cite{CLEO1,CLEO2}. 
We may notice from Table \ref{tab:tab3} that the 
precision obtained by using $e^+e^-$ data alone
has become comparable to the one achieved upon including the 
$\tau$ data. However, one of the latest analyses
reveals a troubling discrepancy between the purely $e^+e^-$ and 
the $\tau$ based
evaluations. Additional work is certainly
needed in order to resolve these problems~\footnote{For
a possible explanation, see \cite{Nyffeler03}.}.
Further data are also
expected in the future, from the KLOE experiment at the DAPHNE
$e^+e^-$ machine \cite{Denig02}, or from the $B$ factories
BaBar \cite{Solodov01} and Belle. For additional comparative discussions 
and details of the
various analyses, we refer the reader
to the literature quoted in Table \ref{tab:tab3}.
Averaging the results from the three most recent
analysis with $e^+e^-$ data only, from Refs. \cite{Jegerlehner01}, 
\cite{Davieretal02}, and \cite{Hagiwaraetal02}, gives \cite{Nyffeler03}
\be
a_{\mu}^{({\mbox{\scriptsize{h.v.p. 1}}})}(e^+e^-)
\,=\, 6838(75) \times 10^{-11}\,.
\lbl{vpaverage}
\ee

\begin{table}[!h]
\caption{Some of the recent evaluations of 
$a_{\mu}^{({\mbox{\scriptsize{h.v.p. 1}}})}\times 10^{11}$
from $e^+e^-$ and/or $\tau$-decay data. 
Note that the authors of Ref. \cite{deTroconizYndurain02} also use 
space like data for the pion form factor $F_{\pi}(t)$; furthermore, 
they use a preliminary version of the CMD-2 data, that is now superseded
by \cite{CMD-2}.}
\begin{center}
\renewcommand{\arraystretch}{1.1}
\begin{tabular}{lcl}
\hline\hline
7024(153) & \cite{EidelmanJegerlehner95} & $e^+e^-$ \\
7026(160) & \cite{BrownWorstell96} & $e^+e^-$   \\
6950(150) & \cite{Alemanyetal98} & $e^+e^-$  \\
7011(94)  & \cite{Alemanyetal98} & $\tau$, $e^+e^-$,   \\
6951(75)  & \cite{DavierHoecker98a} & $\tau$, $e^+e^-$, QCD \\
6924(62)  & \cite{DavierHoecker98b} & $\tau$, $e^+e^-$, QCD \\
7016(119) & \cite{Narison01} &     $e^+e^-$, QCD \\
7036(76)  & \cite{Narison01} &  $\tau$, $e^+e^-$, QCD \\
7002(73)  & \cite{deTroconizYndurain02} & $e^+e^-$, incl. BES-II data, 
                                          $F_{\pi}(t)$ \\
6836(86) & \cite{Jegerlehner01} & $e^+e^-$, incl. BES-II  and CMD-2 data \\
6847(70)  & \cite{Davieretal02} & $e^+e^-$, incl. BES-II and CMD-2 data \\
7090(59)  & \cite{Davieretal02} & $\tau$, $e^+e^-$, incl. BES-II data \\
6831(62)  & \cite{Hagiwaraetal02} & $e^+e^-$, incl. BES-II and CMD-2 data \\
\hline\hline
\end{tabular}
\label{tab:tab3} 
\end{center}
\end{table}

\indent

\noindent
Let us briefly mention here
that it is quite easy to estimate the order of magnitude of 
$a_{\mu}^{({\mbox{\scriptsize{h.v.p. 1}}})}$.
For this purpose, it is convenient to introduce still another 
representation \cite{LautrupdeRafael72,PerrottetdeRafael},
which relates $a_{\mu}^{({\mbox{\scriptsize{h.v.p. 1}}})}$ to the hadronic 
Adler function
${\cal A}(Q^2)$, defined as \footnote{Unlike $\Pi(t)$ itself, ${\cal A}(Q^2)$
if free from ultraviolet divergences.}
\be
{\cal A}(Q^2) \,=\, -Q^2\,\frac{\partial\Pi(Q^2)}{\partial Q^2}\,=\,\int_0^\infty
dt\,\frac{Q^2}{(t + Q^2)^2}\,\frac{1}{\pi}\,{\mbox{Im}}\Pi(t)
\,,
\ee
by
\be
a_{\mu}^{({\mbox{\scriptsize{h.v.p. 1}}})} \,=\,2\pi^2\,\left(\frac{\alpha}{\pi}\right)^2
\int_0^1 \frac{dx}{x} (1-x)(2-x){\cal A}\left(\frac{x^2}{1-x}\,m_\mu^2\right)
\,.
\lbl{a_adler}
\ee
A simple representation of the hadronic Adler function can be obtained
if one assumes that Im$\Pi(t)$ is given by a single, zero width, vector meson pole,
and, above a certain threshold $s_0$, by the QCD perturbative continuum 
contribution,
\be
\frac{1}{\pi}\,{\mbox{Im}}\Pi(t)\,=\, \frac{2}{3}\,f_V^2M_V^2\delta(t-M_V^2)\,+\,
\frac{2}{3}\,\frac{N_C}{12\pi^2}\,\left[1 + {\cal O}(\alpha_s)\right]\,\theta(t-s_0)
\ee
The justification \cite{Perisetal98} for this type of minimal hadronic 
ansatz can be found within the framework
of the large-$N_C$ limit \cite{tHooft74,Witten79} of QCD,
see Ref. \cite{Perisetal98} for a general discussion
and a detailed study of this  representation of the Adler 
function. 
The threshold $s_0$ for the onset of the 
continuum can be fixed from the property that in QCD 
there is no contribution in $1/Q^2$ in the  short distance 
expansion of ${\cal A}(Q^2)$, 
which requires  \cite{Perisetal98}
\be 
2f_V^2M_V^2 \,=\, \frac{N_C}{12\pi^2}\,s_0\,\left(1 \,+\,
\frac{3}{8}\,\frac{\alpha_s(s_0)}{\pi}\,+\, {\cal O}(\alpha_s^2)\right)
\,.
\ee
This then gives 
$a_{\mu}^{({\mbox{\scriptsize{h.v.p. 1}}})}\sim (570 \pm 170)\times 10^{-10}$,
which compares well with the more elaborate data based evaluations
in Table \ref{tab:tab3}, even though this simple estimate cannot 
claim to provide the required accuracy of about 1\%.

\indent

\begin{figure}[!h]
\centerline{\psfig{figure=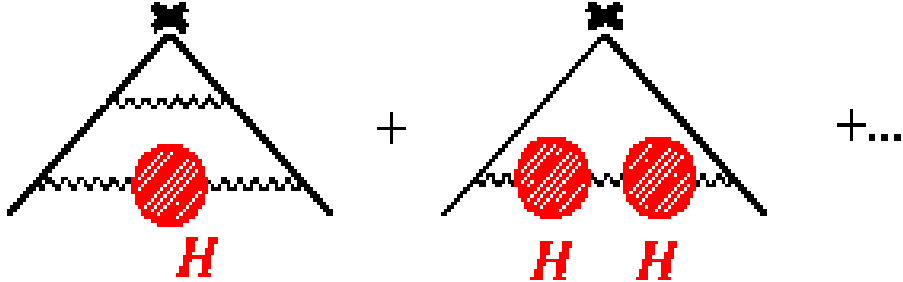,height=4cm,width=12cm}}
\caption{Higher order corrections containing 
the hadronic vacuum polarization contribution, 
corresponding to
$a_{\mu}^{({\mbox{\scriptsize{h.v.p. 2}}})}$.
} \label{fig:amuhadr4}
\end{figure}

\indent

\noindent
We now come to the ${\cal O}(\alpha^3)$ corrections involving
hadronic vacuum polarization subgraphs. Besides the  contributions
shown in Fig.~\ref{fig:amuhadr4}, another one is obtained
upon inserting a lepton loop in one of the two photon lines
of the graph shown in Fig.~\ref{fig:amuhadrp}. Taken all together,
these can
again be expressed in terms of $R^{\mbox{\scriptsize{had}}}$
\cite{Calmetetal76,CalmetetalRMP77,Krause97}
\be
a_{\mu}^{({\mbox{\scriptsize{h.v.p. 2}}})} \,=\,
\frac{1}{3}\,\left(\frac{\alpha}{\pi}\right)^3\,
\int_{4M_\pi^2}^{\infty} \,\frac{ds}{s}\,K^{(2)}(s)R^{\mbox{\scriptsize{had}}}(s)
\,.
\ee
The value obtained
for this quantity is \cite{Krause97}
\be
a_{\mu}^{({\mbox{\scriptsize{h.v.p. 2}}})}\times 10^{11} = -101\pm 6
\,.
\lbl{krause}
\ee
The expression of $K^{(2)}(s)$ is given in the references quoted above.

\indent

\begin{figure}[!h]
\centerline{\psfig{figure=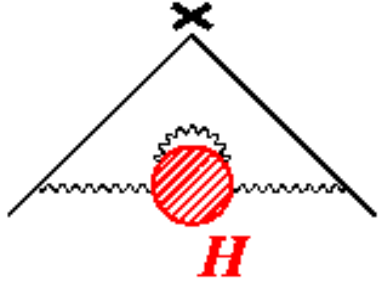,height=4cm,width=5cm}}
\caption{A higher order correction containing 
the hadronic vacuum polarization contribution, 
and which is included in 
$a_{\mu}^{({\mbox{\scriptsize{h.v.p. 1}}})}$.
} \label{fig:amuhadr3}
\end{figure}

\indent

\noindent
There is actually another ${\cal O}(\alpha^3)$ correction,
namely the one obtained upon attaching a virtual photon
line with both ends to the hadronic blob in 
Fig.~\ref{fig:amuhadrp}, see Fig.~\ref{fig:amuhadr3}. On the other hand, 
$a_{\mu}^{({\mbox{\scriptsize{h.v.p. 1}}})}$ involves
in principle data corrected for {\it all} electromagnetic
effects. Whereas radiative corrections in the leptonic
initial state and vacuum polarization effects in
the photon propagator certainly can be accounted for, there is at
present no way to handle in a model independent way electromagnetic
corrections in the hadronic final state. These, on the other hand,
contribute, together with final states containing an additional photon,
to the ${\cal O}(\alpha^3)$ contribution
we have just been mentioning. It has become customary to include
all these effects into $a_{\mu}^{({\mbox{\scriptsize{h.v.p. 1}}})}$,
where it is then to be understood that $R^{\mbox{\scriptsize{had}}}(s)$
in Eq. \rf{hvp1} actually stands for
\be
R^{\mbox{\scriptsize{had}}}(s) \,=\,
\frac{\sigma_0^{e^+e^-\to\mbox{\scriptsize{had}}}(s)\,+\,
\sigma_2^{e^+e^-\to\mbox{\scriptsize{had}}}(s)\,+\,
\sigma_0^{e^+e^-\to
\mbox{\scriptsize{had}}+\gamma}(s)}{\sigma_\infty^{e^+e^-\to\mu^+\mu^-}(s)}
\,,
\lbl{Rhadtot}
\ee
where $\sigma_0^{e^+e^-\to\mbox{\scriptsize{had}}}(s)\,+\,
\sigma_2^{e^+e^-\to\mbox{\scriptsize{had}}}(s)$
denotes the cross section for $e^+e^-\to\mbox{{had}}$ beyond
leading order in the expansion in powers of the fine
structure constant $\alpha$.
The values given in Table \ref{tab:tab3} correspond to
the definition \rf{Rhadtot}.
It also should be stressed that the next-to-leading
cross section $\sigma_2^{e^+e^-\to\mbox{\scriptsize{had}}}(s)$,
as well as the radiative cross section
$\sigma_0^{e^+e^-\to \mbox{\scriptsize{had}}+\gamma}(s)$,
are  infrared divergent quantities, and only their
sum is actually well defined.

\subsubsection{Hadronic light-by-light scattering}

\begin{figure}[!h]
\centerline{\psfig{figure=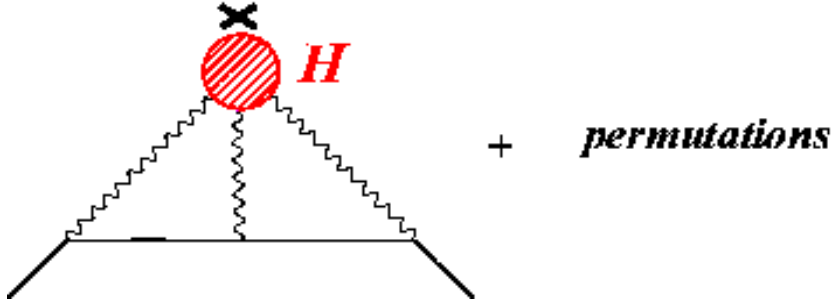,height=4cm,width=11cm}}
\caption{The hadronic light-by-light
scattering graphs contributing to
$a_{\mu}^{({\mbox{\scriptsize{h. L$\times$L}}})}$.
} \label{fig:hadrll}
\end{figure}

We now discuss the so called hadronic light-by-light scattering
graphs of Fig. \ref{fig:hadrll}. 

\indent

\noindent

\begin{figure}[!h]
\centerline{\psfig{figure=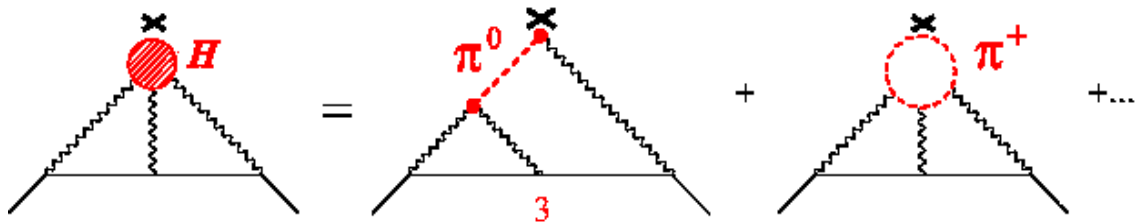,height=3cm,width=15cm}}
\caption{Some individual contributions to hadronic
light-by-light scattering: the neutral 
pion exchange and the charged pion loop. There
are other contributions, not shown here.
} \label{fig:WZll}
\end{figure}

\noindent
The contribution to $\Gamma_{\rho}(p\,',p)$ of relevance here is the 
matrix element, 
at lowest nonvanishing order in the fine structure constant 
$\alpha$, 
of the light quark electromagnetic current
\be
j_{\rho}(x) \,=\, \frac{2}{3}({\bar u}\gamma_{\rho}u)(x)\,-
\,\frac{1}{3}({\bar d}\gamma_{\rho} d)(x)
\,-\,\frac{1}{3}({\bar s}\gamma_{\rho} s)(x)
\lbl{current}
\ee
between $\mu^-$ states,
\bea
(-ie){\bar{\mbox{u}}}(p\,')
{\Gamma}^{({\mbox{\scriptsize{h. L$\times$L}}})}_{\rho}(p\,',p)\mbox{u}(p) &\equiv &
\langle \mu^-(p\,')\vert (ie)j_{\rho}(0) \vert \mu^-(p)\rangle
\nonumber\\
&=&
\int\frac{d^4q_1}{(2\pi)^4}\int\frac{d^4q_2}{(2\pi)^4}\,
\frac{(-i)^3}{q_1^2\,q_2^2\,(q_1+q_2-k)^2}\,
\nonumber\\
&
\times &\!\!\!\!\!
\frac{i}{(p\,'-q_1)^2-m^2}\,
\frac{i}{(p\,'-q_1-q_2)^2-m^2}
\nonumber\\
&
\times &\!\!\!\!\!
(-ie)^3\overline{\mbox{u}}(p\,')
\gamma^{\mu}(\not\! p\,'- \not\!q_1+m)
\gamma^{\nu}(\not\! p\,'- \not\! q_1-\not\! q_2+m)
\gamma^{\lambda}\mbox{u}(p)
\nonumber\\
&
\times &\!\!\!\!\!
(ie)^4\Pi_{\mu\nu\lambda\rho}(q_1,q_2,k-q_1-q_2)\,,
\eea
with $k_{\mu}=(p\,' - p)_{\mu}$ and 
\bea
\Pi_{\mu\nu\lambda\rho}(q_1,q_2,q_3) &=&
\int d^4x_1\int d^4x_2\int d^4x_3
\,e^{i(q_1\cdot x_1 + q_2\cdot x_2 + q_3\cdot x_3)}\,
\nonumber\\
&&\quad
\times
\langle\,\Omega\,\vert\,\mbox{T}\{j_{\mu}(x_1)j_{\nu}(x_2)j_{\lambda}(x_3)j_{\rho}(0)\}
\,\vert\,\Omega\,\rangle
\eea
the fourth-rank light quark hadronic vacuum-polarization tensor, $\vert\,\Omega\,\rangle$ 
denoting the QCD vacuum.
%
%
Since the flavour diagonal current $j_{\mu}(x)$ is conserved, the tensor 
$\Pi_{\mu\nu\lambda\rho}(q_1,q_2,q_3)$ satisfies the Ward identities
\be
\{q_1^{\mu};q_2^{\nu};q_3^{\lambda};(q_1+q_2+q_3)^{\rho}\}
\Pi_{\mu\nu\lambda\rho}(q_1,q_2,q_3)\,=\,\{0;0;0;0\}
\,.
\ee
This entails that \footnote{We use the following conventions for Dirac's $\gamma$-matrices: 
$\{\gamma_{\mu},\gamma_{\nu}\}=2\eta_{\mu\nu}$, with $\eta_{\mu\nu}$  the flat Minkowski 
space metric of signature $(+- - -)$, $\sigma_{\mu\nu}=(i/2)[\gamma_{\mu},\gamma_{\nu}]$, 
$\gamma_5=i\gamma^0\gamma^1\gamma^2\gamma^3$, whereas the totally antisymmetric tensor 
$\varepsilon_{\mu\nu\rho\sigma}$ is chosen such that $\varepsilon_{0123}=+1$. }
\be
{\bar{\mbox{u}}}(p\,')
{\Gamma}^{({\mbox{\scriptsize{h. L$\times$L}}})}_{\rho}(p\,',p)\mbox{u}(p)\,=\,
{\bar{\mbox{u}}}(p\,')\bigg[ \gamma_{\rho}
{F}^{({\mbox{\scriptsize{h. L$\times$L}}})}_1(k^2)\,+\,
\frac{i}{2m}\,\sigma_{\rho\tau}k^{\tau}
{F}^{({\mbox{\scriptsize{h. L$\times$L}}})}_2(k^2)\bigg]\mbox{u}(p)
\,,
\ee
as well as 
${\Gamma}^{({\mbox{\scriptsize{h. L$\times$L}}})}_{\rho}(p\,',p)=
k^{\tau}{\Gamma}^{({\mbox{\scriptsize{h. L$\times$L}}})}_{\rho\tau}(p\,',p)$ 
with
\bea
{\bar{\mbox{u}}}(p\,')
{\Gamma}^{({\mbox{\scriptsize{h. L$\times$L}}})}_{\rho\sigma}(p\,',p)\mbox{u}(p) 
&=&
-ie^6\,\int\frac{d^4q_1}{(2\pi)^4}\int\frac{d^4q_2}{(2\pi)^4}\,
\frac{1}{q_1^2\,q_2^2\,(q_1+q_2-k)^2}\,
\nonumber\\
&&\quad
\times
\frac{1}{(p\,'-q_1)^2-m^2}\,
\frac{1}{(p\,'-q_1-q_2)^2-m^2}
\nonumber\\
&&\quad
\times
{\bar{\mbox{u}}}(p\,')
\gamma^{\mu}(\not\! p\,'- \not\!q_1+m)
\gamma^{\nu}(\not\! p\,'- \not\! q_1-\not\! q_2+m)
\gamma^{\lambda}\mbox{u}(p)
\nonumber\\
&&\quad
\times
\frac{\partial}{\partial k^{\rho}}\,
\Pi_{\mu\nu\lambda\sigma}(q_1,q_2,k-q_1-q_2)\,.
\lbl{Gamma2}
\eea
Following Ref. \cite{Aldinsetal70} and using the property
$k^{\rho}k^{\sigma}{\bar{\mbox{u}}}(p\,')
{\Gamma}^{({\mbox{\scriptsize{h. L$\times$L}}})}_{\rho\sigma}
(p\,',p)\mbox{u}(p)=0$, one deduces that
${ F}^{({\mbox{\scriptsize{h. L$\times$L}}})}_1(0)=0$ 
and that the hadronic light-by-light
contribution to the muon anomalous magnetic moment is equal to
\be
a_\mu^{({\mbox{\scriptsize{h. L$\times$L}}})} \,\equiv\,
{F}^{({\mbox{\scriptsize{h. L$\times$L}}})}_2(0)\,=\,
\frac{1}{48m}\,
{\mbox{tr}}\left\{(\not\! p + m)[\gamma^{\rho},\gamma^{\sigma}](\not\! p + m)
{\Gamma}^{({\mbox{\scriptsize{h. L$\times$L}}})}_{\rho\sigma}(p,p)\right\}
\,.
\lbl{F2trace}
\ee
This is about all we can say about the QCD four-point
function $\Pi_{\mu\nu\lambda\rho}(q_1,q_2,q_3)$.
Unlike the hadronic vacuum polarization function, 
there is no experimental data which would allow for
an evaluation of $a_{\mu}^{({\mbox{\scriptsize{h. L$\times$L}}})}$.
The existing estimates regarding this quantity therefore rely
on specific models in order to account for the non perturbative
QCD aspects. A few particular contributions can be identified,
see Fig. \ref{fig:WZll}. For instance,
there is a contribution where the four photon lines are attached
to a closed loop of charged mesons. The case of the charged pion 
loop with pointlike couplings is actually finite and contributes
$\sim 4\times 10^{-10}$ to $a_\mu$ \cite{Kinoshitaetal85}. 
If the coupling of charged pions
to photons is modified by taking into account the effects of
resonances like the $\rho$, this contribution is reduced by a factor
varying between 3 \cite{Kinoshitaetal85, Bijnensetal96}
and 10 \cite{Hayakawaetal96}, depending on the resonance model used.
Another class of contributions consists of those involving resonance exchanges
between photon pairs 
\cite{Kinoshitaetal85,Bijnensetal96,Hayakawaetal96,HayakawaKinoshita98}.
Although here also the results depend on the models used, there is
a constant feature that emerges from all the analyses that have been 
done: the 
contribution coming from the exchange of the pseudoscalars, $\pi^0$,
$\eta$ and $\eta\,'$ gives practically the final result. Other contributions
[charged pion loops, vector, scalar, and axial resonances,...] 
are not only smaller, but also tend to cancel among themselves.

\indent

\indent

\noindent
Some of the results obtained for
$a_{\mu}^{({\mbox{\scriptsize{h. L$\times$L}}})}$
have been collected in Table \ref{tab:tab4}. Leaving aside the
first result \cite{Calmetetal76,CalmetetalRMP77} 
shown there, which is affected by a bad
numerical convergence \cite{Kinoshitaetal85}, one notices
that the sign of this contribution has changed twice. The first
change resulted from a mistake in Ref. \cite{Kinoshitaetal85},
that was corrected for in \cite{Hayakawaetal96}. The minus sign
that resulted was confirmed by an independent calculation, 
using the ENJL model, in Ref. \cite{Bijnensetal96}. A 
subsequent reanalysis \cite{HayakawaKinoshita98} 
gave additional support to a negative result, while also
getting better agreement with the value of
Ref. \cite{Bijnensetal96}.

\begin{table}[!h]
\caption{Various evaluations of 
$a_{\mu}^{({\mbox{\scriptsize{h. L$\times$L}}})}\times 10^{11}$ 
and of the pion pole contribution
$a_{\mu}^{({\mbox{\scriptsize{h. L$\times$L}};\pi^0})}\times 10^{11}$.
}
\begin{center}
\renewcommand{\arraystretch}{1.1}
\begin{tabular}{llr}
\hline\hline
  --260(100)    & constituent quark loop  &  \cite{Calmetetal76,CalmetetalRMP77}    
\\
      +60(4)    & constituent quark loop  &  \cite{Kinoshitaetal85}       
\\
      +49(5)    & $\pi^\pm$loop, $\pi^0$ and resonance exchanges, 
                  $a_{\mu}^{({\mbox{\scriptsize{h. L$\times$L}};\pi^0})}=+65(6)$   
                                                    &  \cite{Kinoshitaetal85}       
\\
 --92(32)       &    ENJL,  
                  $a_{\mu}^{({\mbox{\scriptsize{h. L$\times$L}};\pi^0+\eta+\eta\,'})}=-85(13)$ 
                                                    &\cite{Bijnensetal96}          
\\
     --52(18)   & $\pi^\pm$ loop, $\pi^0$ and resonance exchanges, and quark loop 
                  $a_{\mu}^{({\mbox{\scriptsize{h. L$\times$L}};\pi^0})}=-55.60(3)$ 
                                                    &  \cite{Hayakawaetal96}  \\
 --79.2(15.4)   & $\pi^\pm$ loop, $\pi^0$ pole and quark loop,   
                  $a_{\mu}^{({\mbox{\scriptsize{h. L$\times$L}};\pi^0})}=-55.60(3)$
                                                    &\cite{HayakawaKinoshita98}
\\                    
   +83(12)      & $\pi^0$, $\eta$ and $\eta\,'$ exchanges only   
                                                    &  \cite{KnechtNyffeler02}    
\\
  +89.6(15.4)   & $\pi^\pm$ loop, $\pi^0$ pole and quark loop,   
                  $a_{\mu}^{({\mbox{\scriptsize{h. L$\times$L}};\pi^0})}=+55.60(3)$
                                                    &  \cite{HayakawaKinoshita02} 
                                                     
\\
  +83(32)       &    ENJL,  
                  $a_{\mu}^{({\mbox{\scriptsize{h. L$\times$L}};\pi^0+\eta+\eta\,'})}=+85(13)$ 
                                                    &  \cite{Bijnensetal02}        
\\
\hline\hline
\end{tabular}
\label{tab:tab4} 
\end{center}
\end{table}

\indent

\noindent
Needless to say, these evaluations are based on
heavy numerical work, which has the drawback of
making the final results rather opaque to an 
intuitive understanding of the
physics behind them. We~\cite{KnechtNyffeler02} 
therefore decided to 
improve things on the analytical side, in order to
achieve a better understanding of the relevant
features that led to the previous results. Taking
advantage of the observation that the pion exchange
contribution 
$a_{\mu}^{({\mbox{\scriptsize{h. L$\times$L}};\pi^0})}$
was found to dominate the final values obtained
for $a_{\mu}^{({\mbox{\scriptsize{h. L$\times$L}}})}$,
we concentrated our efforts on that part, that I shall
now describe in greater detail. For a detailed account
on how the other contributions to
$a_{\mu}^{({\mbox{\scriptsize{h. L$\times$L}}})}$
arise, I refer the reader to the original works
\cite{Kinoshitaetal85,Bijnensetal96,Hayakawaetal96,HayakawaKinoshita98}.

\indent

\begin{figure}[!h]
\centerline{\psfig{figure=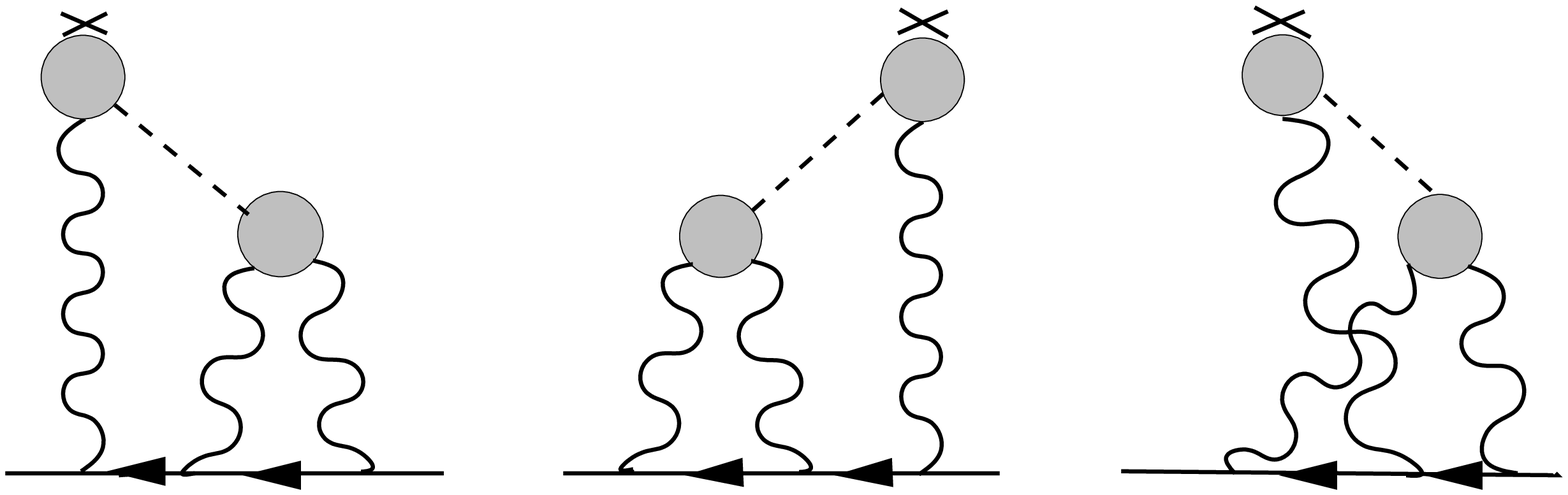,height=4.8cm,width=15cm}}
\caption{The pion-pole contributions to light-by-light scattering. The shaded 
blobs represent the form factor $\FF$. The first and second graphs
give rise to identical contributions, involving the function
$T_1(q_1,q_2;p)$ in Eq. \protect\rf{a_pion_2}, whereas the third graph
gives the contribution involving $T_2(q_1,q_2;p)$.
} \label{fig:pionpole}
\end{figure}

\noindent
The contributions to 
$\Pi_{\mu\nu\lambda\rho}(q_1,q_2,q_3)$ due to single 
neutral pion exchanges, see Fig. \ref{fig:pionpole}, read
\bea
\Pi_{\mu\nu\lambda\rho}^{(\pi^0)}(q_1,q_2,q_3) & = & i
\,{\FF(q_1^2, q_2^2) \ \FF(q_3^2, (q_1+q_2+q_3)^2) \over (q_1+q_2)^2 - 
M_{\pi}^2} \ \varepsilon_{\mu\nu\alpha\beta} \, q_1^\alpha q_2^\beta \ 
\varepsilon_{\lambda\rho\sigma\tau} \, q_3^\sigma (q_1 + q_2)^\tau
\nonumber \\ 
& &
\!\! \,+ i\, {\FF(q_1^2, (q_1 + q_2 + q_3)^2) \ \FF(q_2^2, q_3^2) \over
(q_2+q_3)^2 - M_{\pi}^2} \ \varepsilon_{\mu\rho\alpha\beta} \,
q_1^\alpha (q_2 + q_3)^\beta \ \varepsilon_{\nu\lambda\sigma\tau} \,
q_2^\sigma q_3^\tau \nonumber \\
&&
\!\! +\, i\, {\FF(q_1^2, q_3^2) \ \FF(q_2^2, (q_1+q_2+q_3)^2) \over
(q_1+q_3)^2 - M_{\pi}^2} \ \varepsilon_{\mu\lambda\alpha\beta} \,
q_1^\alpha q_3^\beta \  \varepsilon_{\nu\rho\sigma\tau} \, q_2^\sigma
(q_1 + q_3)^\tau \ . \nonumber \\ 
& & 
\lbl{Pipionpole}
\eea
The form factor $\FF(q_1^2, q_2^2)$, which corresponds
to the shaded blobs in Fig. \ref{fig:pionpole}, is defined as 
\be
i \int d^4 x e^{i q \cdot x} \langle\,\Omega | T \{ j_\mu(x) j_\nu(0)
\} | \pi^0(p) \rangle \,=\, \varepsilon_{\mu\nu\alpha\beta} \, q^\alpha
p^\beta \, \FF(q^2,(p-q)^2) \, , 
\ee 
with $\FF(q_1^2,q_2^2) = \FF(q_2^2,q_1^2)$.
Inserting the expression \rf{Pipionpole} into \rf{Gamma2} and computing the 
corresponding Dirac traces in Eq. \rf{F2trace}, we obtain
\bea
a_{\mu}^{({\mbox{\scriptsize{h. L$\times$L}};\pi^0})}& = & e^6 
\int {d^4 q_1 \over (2\pi)^4} \int {d^4 q_2 \over (2\pi)^4} 
\,\frac{1}{q_1^2 q_2^2 (q_1 + q_2)^2[(p+ q_1)^2 - m^2][(p - q_2)^2 - m^2]} 
\nonumber \\
&& \quad \quad \times \left[ 
{\FF(q_1^2, (q_1 + q_2)^2) \ \FF( q_2^2, 0) \over q_2^2 - 
M_{\pi}^2} \ T_1(q_1,q_2;p) \nonumber \right. \\ 
&& \quad \quad \quad + \left. {\FF( q_1^2,  q_2^2) \ \FF( (q_1 + q_2)^2,
0) \over (q_1 + q_2)^2 - M_{\pi}^2} \ T_2(q_1,q_2;p) \right] \,, 
\lbl{a_pion_2}  
\eea
where  $T_1(q_1,q_2;p)$ and $T_2(q_1,q_2;p)$ denote two
polynomials in the invariants $p\cdot q_1$, $p\cdot q_2$, $q_1\cdot q_2$.
Their expressions can be found in Ref.~\cite{KnechtNyffeler02}. 
The former arises from the two first diagrams shown in 
Fig. \ref{fig:pionpole}, which give identical contributions, while
the latter corresponds to the third diagram on this same figure.
At this stage, it should also be pointed out that the expression 
\rf{Pipionpole} does not, strictly 
speaking, represent the contribution arising from the pion pole only.
The latter would require that the numerators in \rf{Pipionpole} 
be evaluated at the values of the momenta that correspond to the 
pole indicated by the corresponding denominators. For instance, the 
numerator of the term
proportional to $T_1(q_1,q_2;p)$ in Eq. \rf{a_pion_2}
should rather read $\FF(q_1^2, (q_1^2 + 2q_1\cdot q_2 + M_\pi^2) \ \FF( M_\pi^2, 0)$
with $q_2^2=M_\pi^2$. However, Eq. \rf{a_pion_2} corresponds to what previous authors
have called the pion pole contribution, and which they had found
to dominate the final result. 

\indent

\noindent
{F}rom here on, information on the form factor $\FF(q_1^2,q_2^2)$ is
required in order to proceed. The simplest model for the form factor 
follows from the Wess-Zumino-Witten (WZW) term~\cite{WZW1,WZW2} that 
describes the
Adler-Bell-Jackiw anomaly~\cite{ABJ1,ABJ2} in chiral perturbation
theory. Since in this case the form factor is constant, one needs an
ultraviolet cutoff, at least in the contribution to Eq.~\rf{a_pion_2}
involving $T_1$, the one involving $T_2$ gives a finite
result even for a constant form factor \cite{Kinoshitaetal85}. 
Therefore, this model cannot be used for a reliable
estimate, but at best serves only illustrative purposes in the
present context.\footnote{In the context of an effective field 
theory approach, the pion pole with WZW vertices represents a
chirally suppressed, but large-$N_C$ dominant contribution, whereas
the charged pion loop is dominant in the chiral
expansion, but suppressed in the large-$N_C$ limit 
\cite{deRafael94}.}
Previous calculations \cite{Kinoshitaetal85,Hayakawaetal96,HayakawaKinoshita98}
have also used the usual vector meson dominance
form factor [see also Ref.~\cite{BijnensPersson01}]. 
The expressions for the form factor $\FF$ based on the ENJL
model that have been used in Ref.~\cite{Bijnensetal96} do not allow a
straightforward analytical calculation of the loop integrals. However,
compared with the results obtained in 
Refs.~\cite{Kinoshitaetal85,Hayakawaetal96,HayakawaKinoshita98},
the corresponding numerical estimates are rather close to the VMD case
[within the error attributed to the model dependence]. 
Finally,  representations of the form
factor $\FF$, based on the large-$N_C$ approximation to QCD and that
takes into account constraints from chiral symmetry at low energies,
and from the operator product expansion at short distances, have
been discussed in Ref. \cite{KnechtNyffeler01} . 
They involve either one vector
resonance [lowest meson dominance, LMD] or two vector resonances
[LMD+V], see \cite{KnechtNyffeler01} for details. 
The  four types of form factors just mentioned
can be written in the form [$F_\pi$ is the pion decay constant]
\be \lbl{FF_f_g}
\FF(q_1^2, q_2^2) \,=\, \frac{F_\pi}{3}\,\bigg[ f(q_1^2)\,-\, 
\sum_{M_{V_i}} {1 \over q_2^2 - M_{V_i}^2}
g_{M_{V_i}}(q_1^2)\bigg] \, . 
\ee
For the VMD and LMD form factors, the sum in
Eq.~\rf{FF_f_g} reduces to a single term, and the corresponding
function is denoted $g_{M_V}(q^2)$. It depends on the mass
$M_V$ of the vector resonance, which will be identified with
the mass of the $\rho$ meson. 
For our present purposes, it is enough to consider only 
these two last cases, along with the constant WZW form factor. 
The corresponding functions $f(q^2)$ and
$g_{M_{V}}(q^2)$ are displayed in
Table~\ref{tab:f_g}.

\begin{table}[!h]
\caption{The functions $f(q^2)$ and $g_{M_V}(q^2)$ of Eq.
\protect\rf{FF_f_g} for the different form factors.
$N_C$ is the number of colours, taken equal to 3, and 
$F_\pi = 92.4$ MeV is the pion decay constant.
Furthermore, $c_V = \frac{N_C}{4\pi^2}\frac{M_V^4}{F_\pi^2}$.}
\label{tab:f_g}
\begin{center}
\renewcommand{\arraystretch}{1.1}
\begin{tabular}{ccc}
\hline\hline
 & $f(q^2)$ & $g_{M_V}(q^2)$ \\[0.1cm]
\hline
\rule[0mm]{0mm}{1cm} 
$WZW$ & $\displaystyle-\,{\frac{N_C}{4\pi^2F_{\pi}^2}}$  & 0
\\[0.4cm]
$VMD$ & 0 & $\displaystyle{\frac{N_C}{4\pi^2F_{\pi}^2}\,
\frac{M_V^4}{q^2-M_V^2}}$ 
\\[0.4cm]
$LMD$ & $\displaystyle{\frac{1}{q^2-M_V^2}}$ & $\displaystyle{-\,
\frac{q^2+M_V^2-c_V}{q^2-M_V^2}}$
\\
\hline\hline
\end{tabular}
\end{center}
\end{table}

\indent

\begin{figure}[!h]
\centerline{\psfig{figure=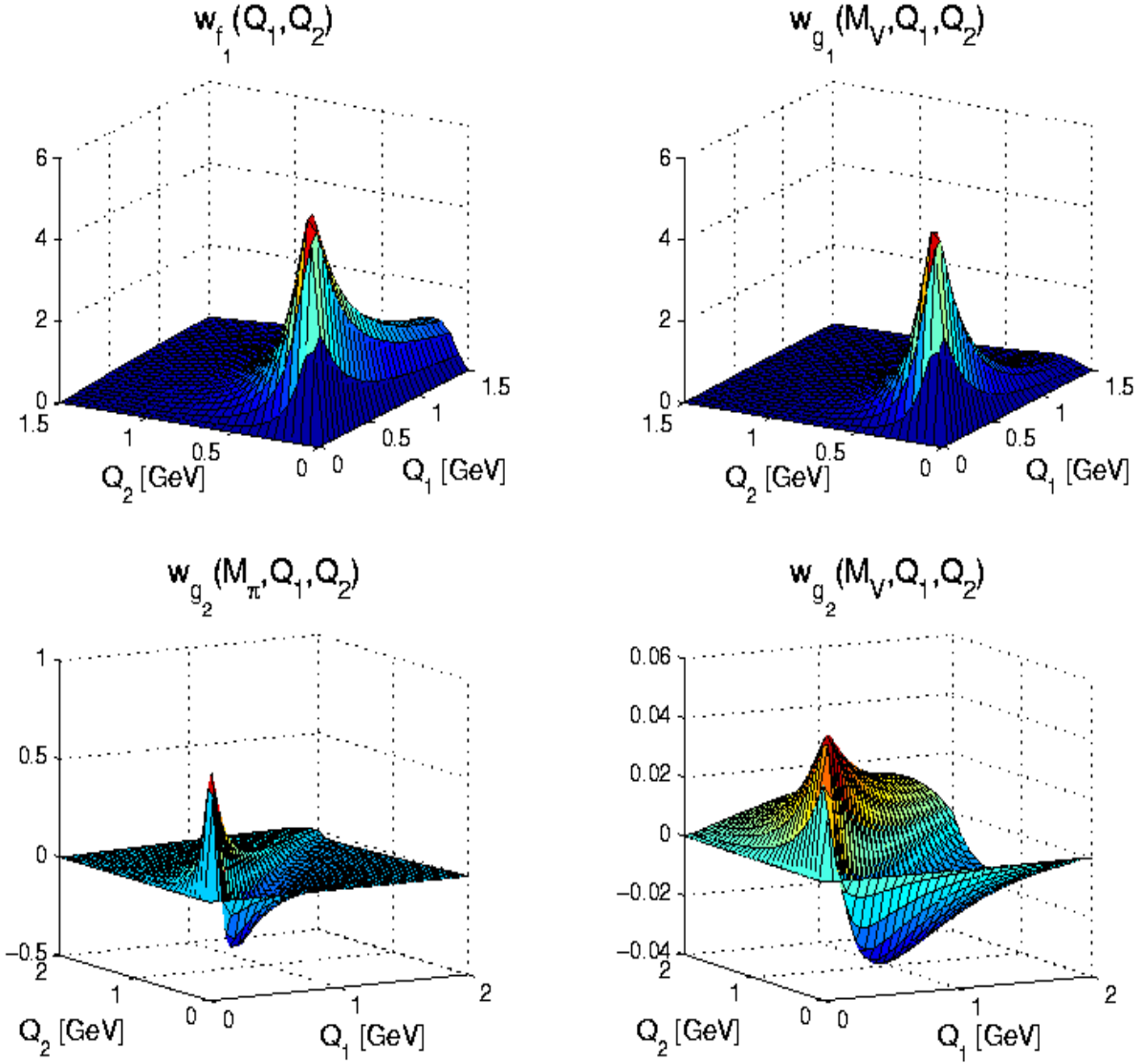,height=15cm,width=16cm}}
\caption{The weight functions appearing in Eqs. \protect{\rf{api1}}
and \protect{\rf{api2}}. Note the different 
ranges of $Q_i$ in the subplots.  The functions $w_{f_1}$ and
$w_{g_1}$ are positive definite and peaked in the region $Q_1\sim
Q_2\sim 0.5$ GeV. Note, however, the tail in $w_{f_1}$ in the
$Q_1$-direction for $Q_2 \sim 0.2~\mbox{GeV}$.  The functions
$w_{g_2}(M_\pi,Q_1,Q_2)$ and $w_{g_2}(M_V,Q_1,Q_2)$ take both signs,
but their magnitudes remain small as compared to $w_{f_1}(Q_1,Q_2)$
and $w_{g_1}(M_V,Q_1,Q_2)$.  We have used $M_V=M_{\rho}=770$ MeV.
} \label{fig:weightfunctions}
\end{figure}

\noindent
We may now come back to Eq. \rf{a_pion_2}. With a representation
of the form \rf{FF_f_g},
the angular integrations can be performed, using for instance
standard Gegenbauer polynomial techniques
[hyperspherical approach], see 
Refs.~\cite{early_Gegenbauer,LRR,RoskiesRemiddiLevineQED}. 
This leads to a two dimensional integral representation:
\bea
a_{\mu}^{({\mbox{\scriptsize{h. L$\times$L}};\pi^0})} & = & 
\left( {\alpha \over \pi } \right)^3 
\left[ a_{\mu}^{(\pi^0;1)} + 
a_{\mu}^{(\pi^0;2)} \right] \, , \lbl{api_two_dim} \\
a_{\mu}^{(\pi^0;1)} 
& = & \int_0^\infty dQ_1 \int_0^\infty dQ_2 \ \Bigg[
w_{f_1}(Q_1,Q_2) \ f^{(1)}(Q_1^2,Q_2^2) \nonumber \\ && \qquad \qquad
\qquad \qquad +  \ w_{g_1}(M_{V},Q_1,Q_2) \
g_{M_{V}}^{(1)}(Q_1^2, Q_2^2) \Bigg] \, , \lbl{api1} \\
a_{\mu}^{(\pi^0;2)} 
& = & \int_0^\infty dQ_1 \int_0^\infty dQ_2 \ \Bigg[
\sum_{M=M_\pi, M_{V}} \ w_{g_2}(M,Q_1,Q_2) \ g_{M}^{(2)}(Q_1^2,
Q_2^2) \Bigg] \, . \lbl{api2}
\eea
The functions $f^{(1)}(Q_1^2,Q_2^2)$, $g_{M_{V}}^{(1)}(Q_1^2,Q_2^2)$,
$g_{M_\pi}^{(2)}(Q_1^2,Q_2^2)$ and $g_{M_{V}}^{(2)}(Q_1^2,Q_2^2)$ are 
expressed in terms of the functions given in Table \ref{tab:f_g},
see Ref. \cite{KnechtNyffeler02}, where
the universal [for the class of form factors that have a representation of the type 
shown in Eq.~\rf{FF_f_g}] weight functions $w_{f_1}$, $w_{g_1}$, and
$w_{g_2}$ in Eqs.~\rf{api1} and
\rf{api2} can also be found. The latter are 
shown in Fig. \ref{fig:weightfunctions}.

\indent

\noindent
The functions $w_{f_1}$ and $w_{g_1}$ are positive and concentrated
around momenta of the order of $0.5~\mbox{GeV}$. This feature was
already observed numerically in Ref.~\cite{Bijnensetal96} by varying the 
upper bound of the
integrals [an analogous analysis is contained in
Ref.~\cite{Hayakawaetal96}].  Note, however, the tail in $w_{f_1}$ in the
$Q_1$ direction for $Q_2 \sim 0.2~\mbox{GeV}$. On the other hand, the
function $w_{g_2}$ has positive and negative contributions in that
region, which will lead to a strong cancellation in the corresponding
integrals, provided they are multiplied by a positive function
composed of the form factors [see the numerical results below].
As can be seen from the plots, and checked analytically, the weight
functions vanish for small momenta. Therefore, the integrals are infrared finite.
The behaviours of the weight functions for large values of
$Q_1$ and/or $Q_2$ can also be worked out analytically. From these,
one can deduce that in the case of the WZW form factor, the
corresponding, divergent, integral for $a_{\mu}^{(\pi^0;1)}$
behaves, as a function of the ultraviolet cut off $\Lambda$,
as $a_{\mu}^{(\pi^0;1)}\sim{\cal C}\ln^2\Lambda$, with
\cite{KnechtNyffeler02}
\be \lbl{coeff_C} 
{\cal C} \,=\, 3 \left( { N_C \over 12 \pi} \right)^2 \left({m_\mu \over F_\pi}
\right)^2 \,=\, 0.0248\, . 
\ee
The log-squared behaviour follows from the general
structure of the integral \rf{api1} for $a_{\mu}^{(\pi^0;1)}$ 
in the case of a constant form factor, as pointed out in \cite{Melnikov01}.
The expression \rf{coeff_C} of the coefficient ${\cal C}$ has 
been derived independently, 
in Ref. \cite{Knechtetal02}, through
a renormalization group argument in the effective theory framework.

\begin{table}[!h]
\caption{Results for the terms 
$a_{\mu}^{(\pi^0;1)}$, 
$a_{\mu}^{(\pi^0;2)}$
and for the pion exchange contribution to the anomalous magnetic
moment 
$a_{\mu}^{\mbox{\scriptsize{(h. L$\times$L;$\pi^0$)}}}$ 
according to Eq.~{\protect\rf{api_two_dim}} for
the different form factors considered. In the WZW model, a cutoff of
$1~\mbox{GeV}$ was used in the first contribution, 
whereas the second term is
ultraviolet finite.}

\begin{center}
\renewcommand{\arraystretch}{1.1}
\begin{tabular}{lr@{.}lr@{.}lr@{.}l}
\hline\hline
Form factor & 
\multicolumn{2}{c}{{$a_{\mu}^{(\pi^0;1)}$}}   
     & \multicolumn{2}{c}{{$a_{\mu}^{(\pi^0;2)}$}}  
     & \multicolumn{2}{c}
       {{$a_{\mu}^{({\mbox{\scriptsize{h. L$\times$L;$\pi^0$}}})}\times 10^{10}$}}
\\ 
\hline  
WZW 	& \hspace*{0.25cm} 0 & 095  	& \hspace*{0.25cm}0 & 0020 &
\hspace*{0.65cm}  12 & 2 \\  
VMD	& 0 & 044 	& 0 & 0013	& 5 & 6 \\ 
LMD 	& 0 & 057 	& 0 & 0014 	& 7 & 3 \\
\hline\hline
\end{tabular}
\label{tab:api_models}  
\end{center}
\end{table}

\indent

\noindent
In the case of the other form factors, 
the integration over $Q_1$ and $Q_2$ is finite and can now be performed
numerically. \footnote{In the case of the VMD form factor, an analytical
result is now also available \cite{Bloklandetal02}.} 
Furthermore, since both the VMD and LMD model tend to the WZW
constant form factor as $M_V\to\infty$, the results for $a_{\mu}^{(\pi^0;1)}$
in these models should scale as ${\cal C}\ln^2M_V^2$ for a large resonance
mass. This has been checked numerically, and the value of the coefficient
${\cal C}$ obtained that way is in perfect agreement with the value
given in Eq. \rf{coeff_C}.
The results of the integration over $Q_1$ and $Q_2$ are
displayed in Table \ref{tab:api_models}. They definitely show a sign
difference when compared to those obtained in Refs. 
\cite{Kinoshitaetal85,Hayakawaetal96,HayakawaKinoshita98,BijnensPersson01},
although in absolute value the numbers agree perfectly.
After the results of Table \ref{tab:api_models} were made public 
\cite{KnechtNyffeler02}, previous authors checked their calculations
and, after some time, 
discovered that they had made a sign mistake at some stage
\cite{HayakawaKinoshita02,Bijnensetal02}. 
The results presented in Table \ref{tab:api_models} 
and in Refs. \cite{KnechtNyffeler02,Knechtetal02} have also
received independent confirmations \cite{BardeendeGouvea,Bloklandetal02}.

\indent

\noindent
The analysis of \cite{KnechtNyffeler02}
leads to the following estimates
\be
a_{\mu}^{({\mbox{\scriptsize{h. L$\times$L;$\pi^0$}}})}\,=\,
5.8(1.0)\times 10^{-10}
\,,
\ee
and
\be
a_{e}^{({\mbox{\scriptsize{h. L$\times$L;$\pi^0$}}})}\,=\,
5.1\times 10^{-14}
\,.
\ee
Taking into account the other contributions computed by previous authors,
and adopting a conservative attitude towards the error to be ascribed to
their model dependences, the total contribution to $a_\mu$ coming
from the hadronic light-by-light scattering diagrams amounts to
\be
a_{\mu}^{({\mbox{\scriptsize{h. L$\times$L}}})}\,=\,
8(4)\times 10^{-10}
\,.
\ee

\indent

\noindent
As a last remark, let me point out that the contribution
depicted in Fig. \ref{fig:amuhadr3} also involves the four
point function $\Pi_{\mu\nu\lambda\rho}(q_1,q_2,q_3)$.
It would be interesting to have an evaluation of this
contribution to $a_{\mu}^{({\mbox{\scriptsize{h.v.p. 1}}})}$
based on the same models that were used to evaluate
$a_{\mu}^{\mbox{\scriptsize{h. L$\times$L}}}$. The 
corresponding contribution arising from the neutral pion 
exchange has been evaluated in Ref. \cite{Bloklandetal02}, 
but it is not obvious that it also dominates the complete
result, since the kinematical configuration is different. 
This evaluation would also allow a direct comparison
with the evaluations of Fig. \ref{fig:amuhadr3} based on data.

\indent

\subsection{Electroweak contributions to $a_\mu$}

\begin{figure}[!h]
\centerline{\psfig{figure=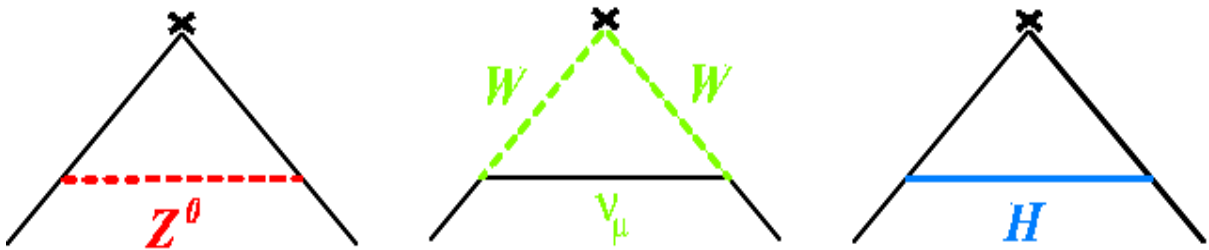,height=3.5cm,width=16cm}}
\caption{One loop weak interaction contributions
to the anomalous magnetic moment.
} \label{fig:W1loop}
\end{figure}

Electroweak corrections to $a_\mu$ have been considered at the
one and two loop levels. The one loop contributions, shown in
Fig. \ref{fig:W1loop}, have been worked out some time ago, and read
\cite{BGL72}-\cite{FLS72}
\be
a_{\mu}^{\mbox{\scriptsize{W(1)}}} \,=\,
\frac{\GF}{\sqrt{2}}\,\frac{m_{\mu}^2}{8\pi^2}\,
\left[\frac{5}{3}+\frac{1}{3}
\left(1-4\sin^2\theta_{W}\right)^2+
\cO\left(\frac{m_{\mu}^2}{M_{Z}^2}\log\frac{M_{Z}^2}{m_{\mu}^2}
\right)+\cO\left(\frac{m_{\mu}^2}{M_{H}^2}\log\frac{M_{H}^2}{m_{\mu}^2}
\right)
\right]\,,
\ee
where the weak mixing angle is defined by
$\sin^2\theta_{W}=1-M_{W}^2/M_{Z}^2$.

\indent

\indent

\noindent
Numerically, with $\GF=1.166
39(1)\times 10^{-5}\,\GeV^{-2}$ and  $\sin^2\theta_{W}=0.224$,
\be
a_{\mu}^{\mbox{\scriptsize
W(1)}}=194.8\times 10^{-11}\,,
\ee

\indent

\noindent
It is convenient to separate the two--loop electroweak contributions into
two sets of Feynman graphs: those which contain closed fermion loops,
which are denoted by $a_{\mu}^{\mbox{\scriptsize{EW(2);f}}}$,
and the others, 
$a_{\mu}^{\mbox{\scriptsize{EW(2);b}}}$. In this notation,
the electroweak contribution to the muon anomalous magnetic moment is
\be
a_{\mu}^{\mbox{\scriptsize{EW}}} \,=\,
a_{\mu}^{\mbox{\scriptsize{W(1)}}}\,+\,
a_{\mu}^{\mbox{\scriptsize{EW(2);f}}}\,+\,
a_{\mu}^{\mbox{\scriptsize{EW(2);b}}}\,.
\ee
I shall review the calculation of the two--loop contributions
separately.

\subsubsection{Two loop bosonic contributions}

The leading logarithmic terms of the two--loop electroweak bosonic
corrections have been extracted using asymptotic expansion
techniques, see e.g. Ref. \cite{Smi94}. In the approximation where
$\sin^2\theta_{W}\ra 0$ and
$M_H\sim M_{W}$ these calculations simplify considerably and one obtains
\be\lbl{LLB}
a_{\mu}^{\mbox{\scriptsize{EW(2);b}}}\,=\,
\frac{\GF}{\sqrt{2}}\,\frac{m_{\mu}^2}{8\pi^2}\,
\frac{\alpha}{\pi}\times 
\left[-\frac{65}{9}\ln\frac{M_{W}^2}{m_{\mu}^2}+
\cO\left(\sin^2\theta_{W}\ln\frac{M_{W}^2}{m_{\mu}^2}
\right)\right]\,.
\ee
In fact, these contributions have now been evaluated analytically, in
a systematic expansion in powers of $\sin^2\theta_{W}$, up to 
$\cO[(\sin^2\theta_{W})^3]\,,$ where $\ln\frac{M_{W}^2}{m_{\mu}^2}$
terms, $\ln\frac{M_{H}^2}{M_{W}^2}$ terms, $\frac{M_{W}^2}{M_{H}^2}
\ln\frac{M_{H}^2}{M_{W}^2}$ terms, $\frac{M_{W}^2}{M_{H}^2}$ terms
and constant terms are kept~\cite{Czarneckietal96}. Using 
$\sin^2\theta_{W}=0.224$ and $M_{H}=250\,\GeV\,,$ the authors of
Ref.~\cite{Czarneckietal96} find
\bea\lbl{bos}
a_{\mu}^{\mbox{\scriptsize{EW(2);b}}} &=&
\frac{\GF}{\sqrt{2}}\,\frac{m_{\mu}^2}{8\pi^2}\,
\frac{\alpha}{\pi}\times 
\left[-5.96\ln\frac{M_{W}^2}{m_{\mu}^2}+0.19\right]
\nonumber\\
&=&
\frac{\GF}{\sqrt{2}}\,\frac{m_{\mu}^2}{8\pi^2}\,
\left(\frac{\alpha}{\pi}\right)\times [-78.9] 
\,=\, -21.1 \times 10^{-11}\,,
\eea
showing, in retrospect, that the simple approximation in
Eq.~\rf{LLB} is rather good.

\indent

\subsubsection{Two loop fermionic contributions}

\begin{figure}[!h]
\centerline{\psfig{figure=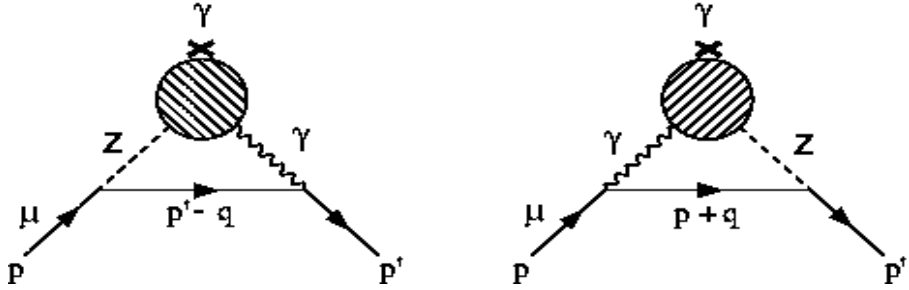,height=4.0cm,width=12cm}}
\caption{Graphs with hadronic contributions to 
 $a_{\mu}^{\mbox{\scriptsize{EW(2);f}}}(\ell,q)$
and involving the QCD three point function
$W_{\mu\nu\rho}(q,k)$.
} \label{fig:W2ggZ}
\end{figure}

The discussion of the two--loop electroweak fermionic corrections is
more delicate. First, it contains a hadronic contribution. Next, 
because of the
cancellation between lepton loops and quark loops in the electroweak
$U(1)$ anomaly, one cannot separate hadronic effects from leptonic
effects any longer. In fact, as discussed in Refs.~\cite{PPdeR95,CKM95},
it is this cancellation which eliminates some of the large logarithms
which, incorrectly were kept in Ref.~\cite{KKSS92}. It is therefore
appropriate to separate the two--loop electroweak fermionic corrections
into two classes: One is the class arising from Feynman diagrams 
containing a lepton or a quark loop, with the external photon, a virtual
photon and a virtual $Z^0$ attached to it, 
see Fig. \ref{fig:W2ggZ}.\footnote{If one works in
a renormalizable gauge, the contributions where the $Z^0$ is replaced
by the neutral unphysical Higgs should also be included. The final
result does not depend on the gauge fixing parameter $\xi_Z$, if one 
works in the class of 't Hooft gauges.} The quark loop of course 
again represents non perturbative hadronic contributions which have to be 
evaluated using some model. This first class is denoted 
by $a_{\mu}^{\mbox{\scriptsize{EW(2);f}}}(\ell;q)$. It involves the QCD
correlation function
\be\lbl{VAV} 
W_{\mu\nu\rho}(q,k)=
\int d^4x\,e^{iq\cdot x}\int d^4y\, e^{i(k-q)\cdot y} \langle
\Omega\,\vert
\mbox{T}\{j_{\mu}(x)A_{\nu}^{(Z)}(y)j_{\rho}(0)\}\vert
\Omega\rangle\,, 
\ee
with $k$ the incoming external photon
four-momentum associated with the classical external magnetic
field. As previously, $j_\rho$ denotes the hadronic part of the
electromagnetic current, and $A_{\rho}^{(Z)}$ is the axial
component of the current which couples the quarks to the $Z^0$
gauge boson.
The other class is defined by the rest of the diagrams, where quark
loops and lepton loops can be treated separately, and is called 
$a_{\mu}^{\mbox{\scriptsize{EW(2);f}}}({\mbox{residual}})$.

\indent

\indent

\noindent
The contribution from 
$a_{\mu}^{\mbox{\scriptsize{EW(2);f}}}({\mbox{residual}})$
brings in factors of the ratio $m_{t}^2/M_{W}^2$. It has been
evaluated, to a very good approximation, in Ref.~\cite{CKM95}, with the
result
\be
a_{\mu}^{\mbox{\scriptsize{EW(2);f}}}({\mbox{residual}})\,=\,
\frac{\GF}{\sqrt{2}}\,\frac{m_{\mu}^2}{8\pi^2}\,
\frac{\alpha}{\pi}\times\left[
\frac{1}{2\sin^2\theta_{W}}\left(-\frac{5}{8}
\frac{m_{t}^2}{M_{W}^2}-\log\frac{m_{t}^2}{M_{W}^2}-\frac{7}{3}\right)
+\Delta_{\mbox{\rm\tiny Higgs}}\right]\,,
\ee
where $\Delta_{\mbox{\rm\tiny Higgs}}$ denotes the contribution from
diagrams with Higgs lines, which the authors of Ref.~\cite{CKM95}
estimate to be 
\be
\Delta_{\mbox{\rm\tiny Higgs}}=-5.5\pm 3.7\,,
\ee
and therefore,
\be\lbl{ferrest}
a_{\mu}^{\mbox{\scriptsize{EW(2);f}}}({\mbox{residual}})\,=\,
\frac{\GF}{\sqrt{2}}\,\frac{m_{\mu}^2}{8\pi^2}\,
\frac{\alpha}{\pi}\times[-21(4)]\,=\,
- 5.6(1.4)\times 10^{-11}\,.
\ee

\indent

\noindent
Let us finally discuss the contributions to
$a_{\mu}^{\mbox{\scriptsize{EW(2);f}}}(\ell;q)$. 
Here, it is convenient to treat
the contributions from the three generations separately. The
contribution from the third generation can be calculated in a
straightforward way, with the result~\cite{PPdeR95,CKM95}
{\setl
\bea\lbl{3rdg}
a_{\mu}^{\mbox{\scriptsize{EW(2);f}}}(\tau;t,b) & = & \frac{\GF}{\sqrt{2}}\,\frac{m_{\mu}^2}
{8\pi^2}\,
\frac{\alpha}{\pi}  \times
 \left[-3\ln\frac{M_{Z}^2}{m_{\tau}^2}-\ln\frac{M_{Z}^2}{m_{b}^2}-
\frac{8}{3}\ln\frac{m_{t}^2}{M_{Z}^2}+\frac{8}{3} +
\cO\left(\frac{M_{Z}^2}{m_{t}^2}\ln\frac{m_{t}^2}{M_{Z}^2}
\right)
\right] \nn \\
 &= & \frac{\GF}{\sqrt{2}}\,\frac{m_{\mu}^2}
{8\pi^2}\,
\frac{\alpha}{\pi}\times [-30.6]
\,=\, -8.2 \times 10^{-11}\,.
\eea}

\noi
In fact the terms of $\cO\left(\frac{M_{Z}^2}{m_{t}^2}
\ln\frac{m_{t}^2}{M_{Z}^2}
\right)$ and  $\cO\left(\frac{M_{Z}^2}{m_{t}^2}\right)$ have also been
calculated in Ref.~\cite{CKM95}. There are in principle QCD
perturbative corrections to this estimate, which have not been
calculated, but the result in Eq.~\rf{3rdg} is good enough for the
accuracy required at present.
The contributions of the remaining charged standard model fermions 
involve the light quarks $u$ and $d$, as well as the second generation
$s$ quark, for which non perturbative effects tied to the spontaneous
breaking of chiral symmetry are important \cite{PPdeR95,KPPdeR02}. 
The contributions  from
the first and second generation are thus most conveniently taken 
together, with the result
{\setl
\bea\lbl{12g}
a_{\mu}^{\mbox{\scriptsize{EW(2);f}}}(e,\mu;u,d,s,c) &=
&\frac{\GF}{\sqrt{2}}\,\frac{m_{\mu}^2} {8\pi^2}\,
\frac{\alpha}{\pi}  \times  \left\{-3\ln
\frac{M_{Z}^2}{m_{\mu}^2} -\frac{5}{2}
\right. \nn \\
& &  \ \ \ \ \ \ \ \ \ \ \ \ \ \ \ \ \ \ \  \
-3\ln\frac{M_{Z}^2} {m_{\mu}^2}+4\ln\frac{M_{Z}^2}{m_{c}^2}
-\frac{11}{6}+\frac{8}{9}\pi^2-8
\nn
\\ &+ &\! \left. 
\left[\frac{4}{3}\ln\frac{M_{Z}^2}{m_{\mu}^2}\!+\!\frac{2}{3}\!+\!\cO\left(
\frac{m_{\mu}^2}{M_{Z}^2}\ln\frac{M_{Z}^2}{m_{\mu}^2}
\right)\!\right]\! +\!4.67(1.80) +0.04(2)\right\}
\nn
\\
& = & \frac{\GF}{\sqrt{2}}\,\frac{m_{\mu}^2} {8\pi^2}\,
\frac{\alpha}{\pi} \times [-28.5(1.8)]\,=\, - 7.6(5) \times 10^{-11}\,,
\eea}
\noindent
where the first line shows
the result from the $e$ loop and the second line the result from the
$\mu$ loop and the $c$ quark, which is treated as a heavy quark. The
term between brackets in the third line is the one induced by the
anomalous term in the hadronic three point function $W_{\mu\nu\rho}(q,k)$
The other contributions have been estimated on the basis of an approximation
to the large-$N_C$ limit of QCD, similar to the one discussed for
the two-point function $\Pi(Q^2)$ after Eq. \rf{a_adler}, see Ref.
\cite{KPPdeR02} for details.

\indent

\noindent
The result in Eq.~\rf{12g} for the
contribution from the first and second generations of quarks and leptons
is conceptually rather different from the corresponding one proposed in
Ref.~\cite{CKM95} ,
{\setl
\bea\lbl{12gCKM}
a_{\mu}^{\mbox{\scriptsize{EW(2);f}}}(e,\mu;u,d,s,c) & = &
\frac{\GF}{\sqrt{2}}\,\frac{m_{\mu}^2} {8\pi^2}\,
\frac{\alpha}{\pi}
\left[-3\ln\frac{M_{Z}^2}
{m_{\mu}^2}+4\ln\frac{M_{Z}^2}{m_{u}^2}-\ln\frac{M_{Z}^2}{m_{d}^2}
-\frac{5}{2}-6
\right. \nn \\
& & \ \ \ \ \ \ \ \ — \  \,\left.
-3\ln\frac{M_{Z}^2}
{m_{\mu}^2}+4\ln\frac{M_{Z}^2}{m_{c}^2}-\ln\frac{M_{Z}^2}{m_{s}^2}
-\frac{11}{6}+\frac{8}{9}\pi^2-6\right] \nn \\
& = & \frac{\GF}{\sqrt{2}}\,\frac{m_{\mu}^2} {8\pi^2}\,
\frac{\alpha}{\pi} \times [-31.9]\,=\, -8.5 \times 10^{-11}
\,,
\eea}
\noindent
where the light quarks were, {\it arbitrarily},  treated the same way as
heavy quarks, with
$m_{u}=m_{d}=0.3\,\GeV\,,$ and $m_{s}=0.5\,\GeV\,.$ 
Numerically, the two expression lead to similar results, though.
A more recent  analysis \cite{Czarneckietal03} provides
a non perturbative treatment of the light quark
sector, and gives 
\be
a_{\mu}^{\mbox{\scriptsize{EW(2);f}}}(e,\mu;u,d,s,c) \, = \,
\frac{\GF}{\sqrt{2}}\,\frac{m_{\mu}^2} {8\pi^2}\,
\frac{\alpha}{\pi} \times [-24.6]\,=\, -6.6 \times 10^{-11}
\,.
\lbl{CMV}
\ee
The difference between the two results in Eqs. \rf{12g}
and \rf{CMV}, which is numerically very small,
is connected to interesting issues
involving the anomalous $\langle VVA\rangle$
three point function in QCD.\footnote{Actually, the
discussion centers around the transverse, i.e. {\it non anomalous}
part of this QCD correlator, and is related to the existence 
of {\it non renormalization theorems}
for it \cite{Vainshtein03,Czarneckietal03,KPPdRinprep}.}

\indent

\noindent
The authors of Ref. \cite{Czarneckietal03} have also performed
a detailed renormalization group analysis of the leading logarithm 
contributions at three loops~\footnote{See also \cite{DG98}.}, and found
them to be negligible. Taking into account other small effects
that were previously neglected, their final value 
reads \cite{Czarneckietal03}
\be
a_{\mu}^{\mbox{\scriptsize{EW}}} \, = \,
15.4(3)\times 10^{-10}\,, 
\ee
which shows  that the two--loop correction represents indeed a reduction
of the one--loop result by an amount of $23\%$. The final error here
includes uncertainties in the hadronic part, the variation of the Higgs mass
in a range 114 GeV $\le M_H \le$ 250 GeV, the uncertainty on the mass of the 
top quark, and unknown three loop effects.

\subsection{Comparison with experiment}

We may now put all the pieces together and obtain the
value for $a_\mu$ predicted by the standard model.
We  have seen that in the case of the hadronic vacuum
polarization contributions, the latest evaluation \cite{Davieretal02}
shows a discrepancy between the value obtained 
exclusively from $e^+e^-$ data and the value that 
arises if $\tau$ data are also included. This gives
us the two possibilities
\bea
a_\mu^{\mbox{\scriptsize{SM}}}(e^+e^-)
&=& (11\,659\,167.5 \pm 7.5 \pm 4.0 \pm 0.4)\times 10^{-10}
\nonumber\\
a_\mu^{\mbox{\scriptsize{SM}}}(\tau)
&=& (11\,659\,192.7 \pm 5.9 \pm 4.0 \pm 0.4)\times 10^{-10}
\,,
\nonumber\\
\eea
where the first error comes from hadronic vacuum polarization, 
the second from hadronic light-by-light scattering, and the last from the
QED and weak corrections. When compared to the present experimental average
\be
a_\mu^{\mbox{\scriptsize{exp}}} \,=\,(11\,659\,203\pm 8)\times 10^{-10}
\ee
there results a difference,
\bea
a_\mu^{\mbox{\scriptsize{exp}}} \,-\, a_\mu^{\mbox{\scriptsize{SM}}}(e^+e^-)
&=&
35.5(11.7) \times 10^{-10}
\,,
\nonumber\\
a_\mu^{\mbox{\scriptsize{exp}}} \, - \, a_\mu^{\mbox{\scriptsize{SM}}}(\tau)
&=&
10.3(10.7) \times 10^{-10}
\,,
\nonumber
\eea
which represents 3.0 and 1.0 standard deviations, respectively.

\indent

\noindent
Although experiment and theory have now both reached the same 
level of accuracy, $\sim\pm 8 \times 10^{-10}$ or $0.7$ ppm,
the present discrepancy between the $e^+e^-$ and $\tau$ based evaluations
makes the interpretation of the above results a delicate issue
as far as evidence for new physics is concerned.
Other evaluations of comparable accuracy 
\cite{DavierHoecker98b,deTroconizYndurain02,Narison01} cover
a similar range of variation in the difference between experiment and
theory. 
Furthermore, the 
value obtained for $a_\mu^{\mbox{\scriptsize{SM}}}(e^+e^-)$ relies 
strongly on the low-energy data obtained by the CMD-2 experiment, with
none of the older data able to check them at the same level of precision.
There seems to be an error in these data from the CMD-2 experiment 
\cite{Nyffeler03}, but this clearly needs to be confirmed.
In this respect, the prospects for additional high statistics data in the 
future, either from KLOE or from BaBar, are most welcome.
On the other hand, if the present discrepancy in the evaluations of the 
hadronic vacuum polarization finds a solution in the future, and if 
the experimental error is further reduced, by, say, a factor of two,
then the theoretical uncertainty on the hadronic light-by-light scattering 
will constitute the next serious limitation on the theoretical side.
It is certainly worthwhile to devote further efforts to a 
better understanding of this contribution, for instance
by finding ways to feed
more constraints with a direct link to QCD into the descriptions
of the four-point function $\Pi_{\mu\nu\rho\sigma}(q_1,q_2,q_3)$.


\indent

\section{Concluding remarks}
\renewcommand{\theequation}{\arabic{section}.\arabic{equation}}
\setcounter{equation}{0}

\noindent
With this review, I hope to have convinced the reader that the subject
of the anomalous magnetic moments of the electron and of the muon 
is an exciting and fascinating topic. It provides a good example 
of mutual stimulation and strong interplay between experiment
and theory.

\indent

\noindent
The anomalous magnetic moment of the 
electron constitutes a very stringent test of QED and 
of the practical working of the framework of perturbatively renormalized 
quantum field theory at higher orders. It tests the validity of QED
at very short distances, and provides at present the best determination
of the fine structure constant.

\indent

\noindent
The anomalous magnetic moment of the
muon represents the best compromise between
sensitivity to new degrees of freedom describing physics beyond the
standard model and experimental feasibility. Important progress has
been achieved on the experimental side during the last couple of years, 
with the results of the E821 collaboration at BNL. The experimental value
of $a_{\mu}$ is now known with an accuracy of 0.7ppm. Hopefully, the 
Brookhaven experiment will be given the opportunity to reach
its initial goal of achieving a measurement at the 0.35 ppm level.

\indent

\noindent
As can be inferred from the examples 
mentioned in this text, 
the subject constitutes, from a theoretical point of view, 
a difficult and error prone topic, due 
to the technical difficulties encountered 
in the higher loop calculations.
The theoretical
predictions have reached a precision comparable 
to the experimental one, but unfortunately there
appears a discrepancy between the most recent evaluations of the hadronic
vacuum polarization according to whether $\tau$ data are considered
or not. Hopefully, this situation will be clarified soon.
Hadronic contributions, especially from vacuum
polarization and from light-by-light scattering, are responsible for 
the bulk part of the final uncertainty in the theoretical value 
$a_{\mu}^{\scriptsize\mbox{SM}}$. Further efforts are needed in 
order to bring these aspects under better control. 

\indent

\indent

\noindent
{\bf Acknowledgements}:
I wish to thank A. Nyffeler, S. Peris, M. Perrottet, and 
E. de Rafael for stimulating and very pleasant collaborations.
Most of the figures appearing in this text were kindly provided  
by M. Perrottet.  A countless number 
of very informative and useful comments was provided 
by A. Nyffeler, M. Perrottet, and  E. de Rafael.
Finally, I wish to thank the organizers
of the 41th edition of the Schladming school for their invitation
to present these lectures and for providing, together with the
students and the other lecturers, a very pleasant and 
fruitful atmosphere.
This work is supported in part by the EC contract 
No. HPRN-CT-2002-00311 (EURIDICE).

\end{document}